\documentclass[fleqn,usenatbib]{mnras}

\usepackage{newtxtext,newtxmath}
\usepackage{nccmath}
\usepackage[flushleft]{threeparttable}
\usepackage{amsmath}

\usepackage[T1]{fontenc}

\DeclareRobustCommand{\VAN}[3]{#2}
\let\VANthebibliography\thebibliography
\def\thebibliography{\DeclareRobustCommand{\VAN}[3]{##3}\VANthebibliography}

\usepackage{graphicx}	
\usepackage{amsmath}	
\usepackage[justification=centering]{caption}
\usepackage{graphicx}
 \usepackage{setspace}
\usepackage{textcomp}     
\usepackage{url}
\usepackage[]{amsmath}
\usepackage{amsmath}
\usepackage{nccmath}
\usepackage{lineno}
\usepackage{siunitx}
\usepackage{geometry}
\usepackage{pdflscape} 
\usepackage{longtable}
\usepackage{float,lscape}
\usepackage{layout}

\newcommand{\cga}{{2016 CG$_{18}$ }}
\newcommand{\cgans}{{2016 CG$_{18}$}}
\newcommand{\eva}{{2016 EV$_{84}$ }}
\newcommand{\evans}{{2016 EV$_{84}$}}
\newcommand{\gea}{{2016 GE$_{1}$ }}
\newcommand{\geans}{{2016 GE$_{1}$}}
\newcommand{\maps}{{Meteoritics $\&$ Planetary Science}}
\newcommand{\psj}{{Planetary Science Journal}}

\title[CFHT small asteroid streak photometry]{Rotation periods and colours of 10-m scale near-Earth asteroids\\from CFHT target of opportunity streak photometry}

\author[B. T. Bolin et al.]{
B. T. Bolin,$^{1}$\thanks{NASA Postdoctoral Program Fellow}$^{,}$\thanks{E-mail: bryce.bolin@nasa.gov (BTB)}
M. Ghosal,$^{2}$
and R. Jedicke$^{2}$
\\
$^{1}$Goddard Space Flight Center, 8800 Greenbelt Road, Greenbelt, MD 20771, USA,\\
$^{2}$Institute for Astronomy, University of Hawai'i, 2680 Woodlawn Dr., Honolulu, HI 96822, USA\\}
\date{Accepted XXX. Received YYY; in original form ZZZ}

\pubyear{2023}

\begin{document}
\label{firstpage}
\pagerange{\pageref{firstpage}--\pageref{lastpage}}
\maketitle

\begin{abstract}
The rotational properties of $\sim$10~m-scale asteroids are poorly understood with only a few measurements. Additionally, collisions or thermal recoil can spin their rotations to periods less than a few seconds obfuscating their study due to the observational cadence imposed by the long read-out times of charge-coupled device imagers. We present a method to measure the rotation periods of 10~m-scale asteroids using the target of opportunity capability of the Canada France Hawaii Telescope and its MegaCam imager by intentionally streaking their detections in single exposures when they are at their brightest. Periodic changes in brightness as small as $\sim$0.05 mag along the streak can be measured as short as a few seconds. Additionally, the streak photometry is taken in multiple g, r, and i filter exposures enabling the measurement of asteroid colours. The streak photometry method was tested on CFHT observations of three 10~m-scale asteroids, \geans, \cgans, and \evans. Our 3 targets are among the smallest known asteroids with measured rotation periods/colours having some of the shortest known rotation periods. We compare our rotation period and taxonomic results with independent data from the literature and discuss applications of the method to future small asteroid observations.
\end{abstract}




\section{Introduction}

Near Earth Asteroids (NEAs) represent a dynamically young sub-population of ejected Main Belt Asteroids (MBAs). They have short lifetimes on the order of millions of years ending in collisions with the inner planets, the Sun, or ejection from the solar system. While the understanding of the rotational properties of km or 100-m scale asteroids has improved \citep[e.g.,][]{Thirouin2016,Szabo2016}, the rotational properties of $\sim$10~m scale asteroids are poorly understood with only a handful of measurements due, in part, to them being orders of magnitude fainter than larger asteroids. Additionally, the Yarkovsky-O'Keefe-Radzievskii-Paddack (YORP) effect, caused by the change in an asteroid's rotation due to the momentum imparted by re-emitted thermal radiation from solar heating, can spin up 10 m-scale NEAs within their dynamical lifetimes so that they have rotation periods of only a few seconds or less \citep[][]{Bottke2006,Vokrouhlicky2015}. 

Studies on the structure of small asteroids have been mostly limited to computer modeling \citep[e.g.,][]{Sanchez2014}.  Some models suggest that small objects should be `strong', perhaps monolithic objects \citep[e.g.,][]{Bottke2005-linking,Bottke2005-fossilized} yet observations of some of the smallest asteroids suggest that they have low-densities \citep[e.g.,][]{Micheli2014-2011MD,Micheli2013-2012LA}. The discovery of NEO 2008 TC$_3$ before its impact with Earth gave the first opportunity to study the structure of a meter-scale asteroid with observational data and meteoritic samples \citep[][]{Jenniskens2010}. The data indicated that 2008 TC$_3$ was a rapid tumbler less than a few meters in diameter and consisted of a heterogeneous mixture of materials implying it was a rubble pile \citep[][]{Kozubal2011}. Other work has shown that small asteroids on the tens of meters scale tend to rotate above the limit allowed by the cohesive strength of rubble piles \citep[][]{Harris2009a,Warner2009ab}.

Information about the structural cohesion of asteroids 10~m or smaller in diameter is less known compared to larger asteroids. There are only 10 asteroids that are 10~m or smaller in diameter with known periods\citep[e.g.,][]{Thirouin2016} and none are rotating faster than the theoretical bursting rate for rubble piles or solid rock \citep[]{Sanchez2014,Bolin2014}. The lack of detection of rapidly-spinning asteroids may be the result of an observational bias. The YORP effect \citep[][]{Bottke2006,Vokrouhlicky2015} can increase the rotation rate of 10~m NEAs to their theoretical bursting limits of $<$10 radians/s in several orders of magnitude less time than their dynamical lifetimes of millions of years \citep[][]{Hirabayashi2015,Nesvorny2023NEO}. 

Additionally, the measurement of the short rotation periods of 10 m-scale asteroids is difficult due to their faintness. Main belt asteroids (MBAs) between 1~m and 10~m in size are too faint to be observed from Earth. Near-Earth asteroids (NEAs) in this size range can reach V$\sim$ 20 or brighter when passing within a few lunar distances of the Earth \citep[][]{Jedicke2016}. However, close-approaching NEAs are difficult to observe due to their high, 
$\sim$10s of \arcsec per second, rates of motion when this close to the Earth, and are only observable for a few days at a time before they become too faint to observe by most facilities or go into solar conjunction \citep[][]{Bolin2020CD3}. Furthermore, the long observational cadence imposed by the typical $\sim$10s of seconds read-out times of charge-coupled device (CCD) imagers makes detection of short rotation periods difficult \citep[][]{Jedicke2015}. The majority of asteroid lightcurve observations use individual charge-coupled device (CCD) camera exposures, focusing on more distant and slower rotating asteroids such as MBAs \citep[][]{McNeill2018, Hanus2018}, Jupiter Trojans \citep[][]{Szabo2017,Ryan2017}, trans-Neptunian objects \citep[][]{Thirouin2019,Whidden2019,Thirouin2022}, and interstellar objects \citep[][]{Bolin2018,Bolin2020HST}. Lightcurve observations of these distant objects use long exposures and read-out modes requiring 10's of seconds resulting in an observation cadence of $\sim$30-60~s.

Only recently, the advent of complementary metal oxide semiconductor (CMOS) cameras \citep[][]{Harding2016} that can provide continuous imaging at a few hz or faster have allowed high time resolution observations of asteroids \citep[][]{Purdum2021,Pomazan2022}. However, the lack of availability of CMOS cameras at most 3~m or larger optical telescope facilities has limited their availability to mostly meter-scale telescopes restricting their use to observing brighter targets \citep[V$<$17][]{Beniyama2022}. An alternative method to obtain high-time resolution photometry of asteroids is to measure lightcurve variations along a trailed asteroid detection. If an asteroid moving 10s of \arcsec/s is observed with sidereal tracking such that the stars remain stationary in the camera, the asteroid will streak forming an elongated PSF in the direction of its motion \citep[][]{Veres2012ab}.

NEAs moving as fast as a few 10s of \arcsec/s can be imaged using non-sidereal tracking at the asteroid's rate of motion resulting in circular PSF asteroid detections while the background stars are trailed \citep[][]{Sharma2023}. If the asteroid is visibly trailed it is possible to measure variations in its brightness along the trail time resolution of the brightness variations is determined by the trail length and the size of the image resolution element provided by the atmospheric seeing or pixel scale. Similar methodology has been used to obtain $\sim$1~ms lightcurve photometry, of stars observed by CCD cameras operated in continuous read mode \citep[][]{Bianco2009, Daniels2023}, and in serendipitous observations of fast-moving asteroids \citep[][]{Clark2023bolide}. In addition to providing high time-resolution lightcurve photometry of asteroids, streak lightcurves can provide colour estimates if taken in a series of images with several different filters since it provides full lightcurve coverage of rapidly rotating asteroids, minimizing the effect that rotational variations will have on the colour measurements \citep[e.g.,][]{Bolin2020CD3}.

In this work we introduce a method that addresses the challenges of determining the colours and rotation periods of rapidly-spinning asteroids by measuring rotation periods and colours from streak asteroid photometry. Our technique gerrymanders the non-sidereal tracking rate to result in the asteroid moving along a CCD column or row at a rate calculated to provide a good signal-to-noise ratio (SNR) and time resolution. Trailed background stars are used to determine the time-resolved photometric calibration.  We tested the technique with observations of several 10~m scale NEAs with the MegaCam instrument on the Canada France Hawaii Telescope \citep[][]{Hartman2006, Gwyn2012} and extracted the photometry, and determined the rotation periods, and colours of our targets with periodogram analysis.

\section{Observations}

We observed three NEAs, \geans, \cgans, and \evans, in 2016 (see Table S1 of \citet[][]{Bolin2023SALsup}. ) with the Queue Service Operations (QSO) TOO  (QSO program 16AH29, PI R. Jedicke) program using the MegaCam instrument mounted at the prime focus of the 3.6~m CFH telescope \citep[][]{Hartman2006}. Our targets had  r~band magnitudes between  17 and 19 and absolute magnitude, H, between 26.7 and 28.5 corresponding to diameters of $\sim$7-15~m assuming an albedo of 0.15 \citep[][]{Harris2002}. The observation of our targets occurred when they were $\sim$0.01~au from the Earth and moving with an angular rate of 0.3 \arcsec/s or faster. Seeing was between $\sim$0.6-0.7 \arcsec and the CFHT SkyProbe \citep[][]{Cuillandre2002} showed minimal variations in attenuation during the time of our observations ($\lesssim$0.01 mag). The observations were taken at airmass $\sim$1.

The TOO CFHT/MegaCam observations were ideal for our program since they needed to occur immediately following the discovery of our targets before they became fainter or went into solar conjunction \citep[][]{Jedicke2016}.  MegaCam provided excellent coverage of our targets and the background star field with its $\sim$1~sq. deg. field of view and 0.187\arcsec\ pixel scale. Our 60~s exposures of\cgans, \evans, and \gea (column 11 of Table S1) were taken in a serial sequence of Sloan Digital Sky Survey (SDSS) g, r and i band filters \citep[][]{Fukugita1996}. Traditional asteroid lightcurve and color observations alternate between filters to minimize the effect of brightness variations due to their rotation \citep[e.g.,][]{Bolin2023NT} but MegaCam's 90~s filter exchange time is prohibitive so filter changes were executed only after 2-3 exposures. We note that it is unnecessary to change filters rapidly for rotation periods much less than the exposure time.

If a telescope's non-sidereal tracking rates match the sky-plane velocity of an asteroid it will result in a circular PSF for the target of width $\theta_c$, typically 0.6 to 0.7\arcsec\ in our observations, with a signal-to-noise ratio (SNR) in an exposure of duration $t_{\mathrm{exp}}$ of $\mathrm{SNR_c}(t_{\mathrm{exp}})$.  We intentionally trailed our target asteroids to measure short term variations in their brightness caused by their rotation.  To ensure that the SNR in each `seeing resolution element' along the trail was sufficient for our photometric purpose the required trail length, $\theta_t$, is given by rearranging Eq.~2 from \citet[][]{Shao2014} et al:
\begin{ceqn}
\begin{equation}
\theta_t \,=\,\theta_c \left (\frac{\mathrm{SNR_c}(t_{\mathrm{exp}})}{\mathrm{SNR_t}}-1 \right )
\end{equation}
\end{ceqn}
where $10\lesssim\mathrm{SNR_t}\lesssim20$ is the SNR within an individual resolution element along the trail.  The net tracking rate is therefore $\dot{\theta} \, = \, \theta_t \, t_{\mathrm{exp}}^{-1}$ to provide the desired trail length $\theta_t$ and the resulting time resolution along the trail, $\delta_t$, is then
\begin{ceqn}
\begin{equation}
\delta_t \, = \, t_{\mathrm{exp}}  \left ({\frac{\mathrm{SNR_c}(t_{\mathrm{exp}})}{\mathrm{SNR_t}}-1} \right )^{-1}.
\end{equation}
\end{ceqn}
We used 60~s exposures with the expectation that the targets's periods would be on the order of tens of seconds and we would typically take about six images, two in each of three different filters. We tried to target objects with $\mathrm{SNR_0}$ of several 100 resulting in $\delta_t$ on the order of seconds. 

The MegaCam CCD chips are 383$\arcsec\times862\arcsec$ on a side, an aspect ratio of $\sim$2.25, so the telescope's tracking rates were adjusted to result in an asteroid trail in the Y direction to reduce the probability that it would cross a gap between the chips. Overscan, bias, flats, dark and fringe corrections were applied, and zeropoints were obtained from the Elixir pipeline \citep[][]{Magnier2004}. We used the Canadian Astronomy Data Centre Solar System Object Image Search tool \citep[][]{Gwyn2012} to predict the chip in which the asteroid's trail would appear (e.g. Fig.~1).  A complete description of the extraction and reduction of the time series asteroid trail photometry is available in Section S1 and example lightcurves extracted for observations of \geans, \cgans, and \eva are provided in Section S2 of \citep[][]{Bolin2023SALsup}.

\begin{figure}\centering
\hspace{0 mm}
\centering
\includegraphics[width=1\linewidth]{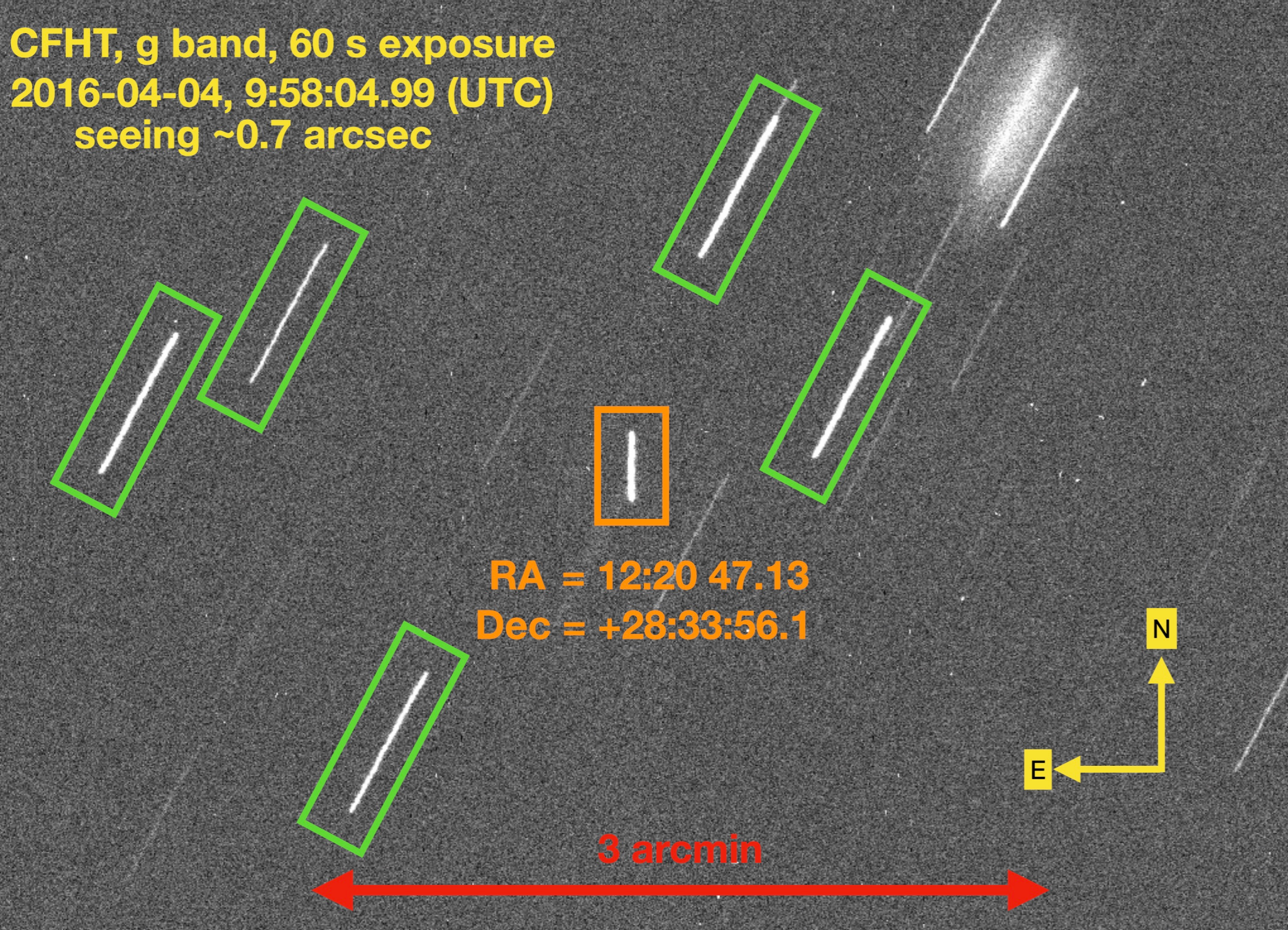}
\caption{Observation of \gea taken in a 60~s g band exposure with CFHT/MegaCam on 2016-Apr-04. The asteroid is outlined in an orange rectangle. The non-sidereial tracking rate of $\sim$0.32 \arcsec/s was adjusted so the asteroid would only trail in the Y direction. This resulted in the reference background stars trailing in a diagonal direction outlined in green rectangles. The cardinal directions and image scale are indicated on the image.}
\end{figure}

We removed secular trends in the lightcurve \citep[e.g.,][]{Lindberg2022} before applying a LS periodogram (LS) periodogram \citep[][]{Lomb1976,Scargle1982} to the combined g, r and i lightcurve data to identify periodicity.  The top panels of Figs.~S5-7 in \citet[][]{Bolin2023SALsup} show the LS periodogram applied to the \geans, \cgans, and \eva g, r and i lightcurve data. The lightcurve period was determined by using the highest peak above the 3-$\sigma$ false alarm probability level, e.g., the peak in the top panel of Fig.~S5 corresponding to a lightcurve period of $\sim$15 s. The data were folded using the double-peaked rotation period and rebinned between a phase of 0 and 1 with bin widths of 0.05 and inspected visually (second and third panels of Figs.~S5-7). The g-r and r-i colour difference as a function of binned phase are shown in the bottom panel of Figs.~S5-7.

\section{Results \& Discussion}

The double-peak rotation period determined from our application of the LS periodogram to the g, r and i data for \geans, \cgans, and \eva are provided in Table~1. Our dataset had a time resolution of $\sim$1~s and the detected rotation periods of our target asteroids are 30-60~s. We applied bootstrap estimation of the uncertainties \citep[][]{Press1986} by randomly removing $\sqrt{N}$ data points from the time-series lightcurves and repeating the periodogram estimation of the lightcurve period 1,000 times resulting in a 1-$\sigma$ estimate of the lightcurve rotation period uncertainties of 0.01-0.1 s. Our lightcurve observations cover multiple full rotations of the targets \citep[second and third panels of Figs.~S5-7 in][]{Bolin2023SALsup}. This is unsurprising given the 750~s to 935~s duration of our lightcurve observations with minimal gaps and the short rotation periods of our targets.

The rotation periods of our three targets are all on the order of tens of seconds, faster than many other observed asteroids in this size range (Fig.~3 and \citet[][]{Thirouin2016,Beniyama2022}), but considerably longer than the predicted spin periods that are on the order of seconds for 10~m-scale asteroids \citep[][]{Farinella1998, Vokrouhlicky2002, Bolin2014}.  The lightcurve amplitudes are $\sim$0.6-0.9 mags (Table~1) providing a rough estimate of the asteroid's b/a axial ratios in the range 1.7-2.3 \citep[][]{Binzel1989}. 

Our measured amplitude of $\sim$0.8~mag for \gea is significantly larger than the $\sim$0.3~mag measured by \citet[][]{Warner2016GE1} on 2016 Apr 5. The difference may be due to the \citet[][]{Warner2016GE1} data having significantly lower SNR ($<$10). Differences in the viewing geometry exacerbated by the close approach at $\sim$0.01~au, e.g., aspect angle which changed from -26 degrees to 35 degrees in the single day separating the two sets of observations, could also cause extreme changes in the observed lightcurve amplitude \citep[e.g.,][]{Barucci1982,Harris2009a}. Additionally, photometry submitted around the same time to the Minor Planet Center showed evidence of a significantly larger lightcurve amplitude than 0.3 mag \citep[][]{Warner2016GE1}.

Our measured rotation period for \gea (30.66$\pm$0.01s) is also significantly shorter than the 33.997$\pm$0.007~s period ($U=2$) reported by \citet{Warner2016GE1}. The difference in the rotation period between the two dates could be due to \gea being in a tumbling state \citep[e.g.,][]{Kaasalainen2001,Pravec2005}.

\begin{table*}
\caption{Rotation periods and colours.}
\centering
\begin{tabular}{lllllll}
\hline
object name    & period & ampltude      & b/a ratio     & m$_r$          & g-r           & r-i           \\
               & (s)    & (mag)         &               & (mag)          & (mag)         & (mag)         \\ \hline
2016 CG$_{18}$ & 55.180$\pm$0.036  & 0.62$\pm$0.07 & 1.77$\pm$0.11 & 17.29$\pm$0.01 & 0.58$\pm$0.01 & 0.17$\pm$0.01 \\
2016 EV$_{84}$ & 52.237$\pm$0.062   & 0.86$\pm$0.16 & 2.21$\pm$0.32 & 18.89$\pm$0.01 & 0.70$\pm$0.02 & 0.05$\pm$0.04 \\
2016 GE$_{1}$  & 30.664$\pm$0.007   & 0.78$\pm$0.05 & 2.05$\pm$0.10 & 17.34$\pm$0.01 & 0.66$\pm$0.01 & 0.20$\pm$0.01 \\ \hline
Solar colours  &        &               &               &                & 0.46$\pm$0.01 & 0.12$\pm$0.01 \\ \hline
\end{tabular}
\begin{tablenotes}
\item \textbf{Notes.} (1) double-peak rotation period, (2) r-band peak-to-peak lightcurve amplitude, (3) Apparent $r$-band magnitude, (4) solar colours from \citet[][]{Haberreiter2017} and \citet[][]{Willmer2018}.
\end{tablenotes}
\end{table*}

The scatter in the g-r and r-i colour measurements for our targets on a bin-by-bin basis is $\sim$0.1 magnitudes which may be due to the large error bars for the colour measurements for individual bins. An additional source of scatter could be due to variations in the object's colour with time as seen for \cgans. We estimated the central values of the colours of our targets by taking the weighted average of the binned, phased colour data. The uncertainties are small because the targets were relatively bright, despite being small objects, because they were rapidly targeted in a relatively long exposure with a 4 meter class telescope while near their brightest during their discovery apparition. The r-i colours span the range from bluer to redder than the Sun (Table~1 and Fig.~2).  The g-r and r-i colours of \gea and \eva overlap the colour-space region generally occupied by S-class asteroids \citep[][]{Ivezic2001} while \cga lies in the border region between the C and S-class. This is as expected since the majority of the small NEA population consists of S-class asteroids \citep[e.g.][]{Jedicke2018} and the source probabilities and albedos of the objects are consistent with being S-class inner main belt objects \citep[Table~2 and ][]{Nesvorny2023NEO,Morbidelli2020}.  

The g-r, r-i colours and spectral slope of \gea (g-r = 0.66$\pm$0.01, r-i = 0.20$\pm$0.01, g-i = 0.86$\pm$0.01, spectral gradient = 9.2$\pm$0.3$\%$) are broadly consistent with being classified as an S-type \citep[S-type spectral gradient = 6-25$\%$][]{DeMeo2013aa}. 

The g-r and r-i colours colours and spectral of \cga (g-r = 0.58$\pm$0.01, r-i = 0.17$\pm$0.01, g-i = 0.75$\pm$0.04, spectral gradient = 6.9$\pm$0.3$\%$) and \eva (g-r = 0.70$\pm$0.02, r-i = 0.05$\pm$0.04, g-i = 0.75$\pm$0.04, spectral gradient = 6.9$\pm$1.3$\%$) are consistent with being classified as an X-class asteroid \citep[X-class spectral gradient = 3-9$\%$,][]{DeMeo2013aa}. Asteroids belonging to both the X and S-type classes are found in the inner-Main belt \citep[][]{DeMeo2013aa}. \citet[][]{Devogele2019} found that 
\cga has a Xe-type spectrum using visible spectroscopy aligning with our results that it has X-complex colours.

\begin{figure}
\centering
\includegraphics[width=1\linewidth]{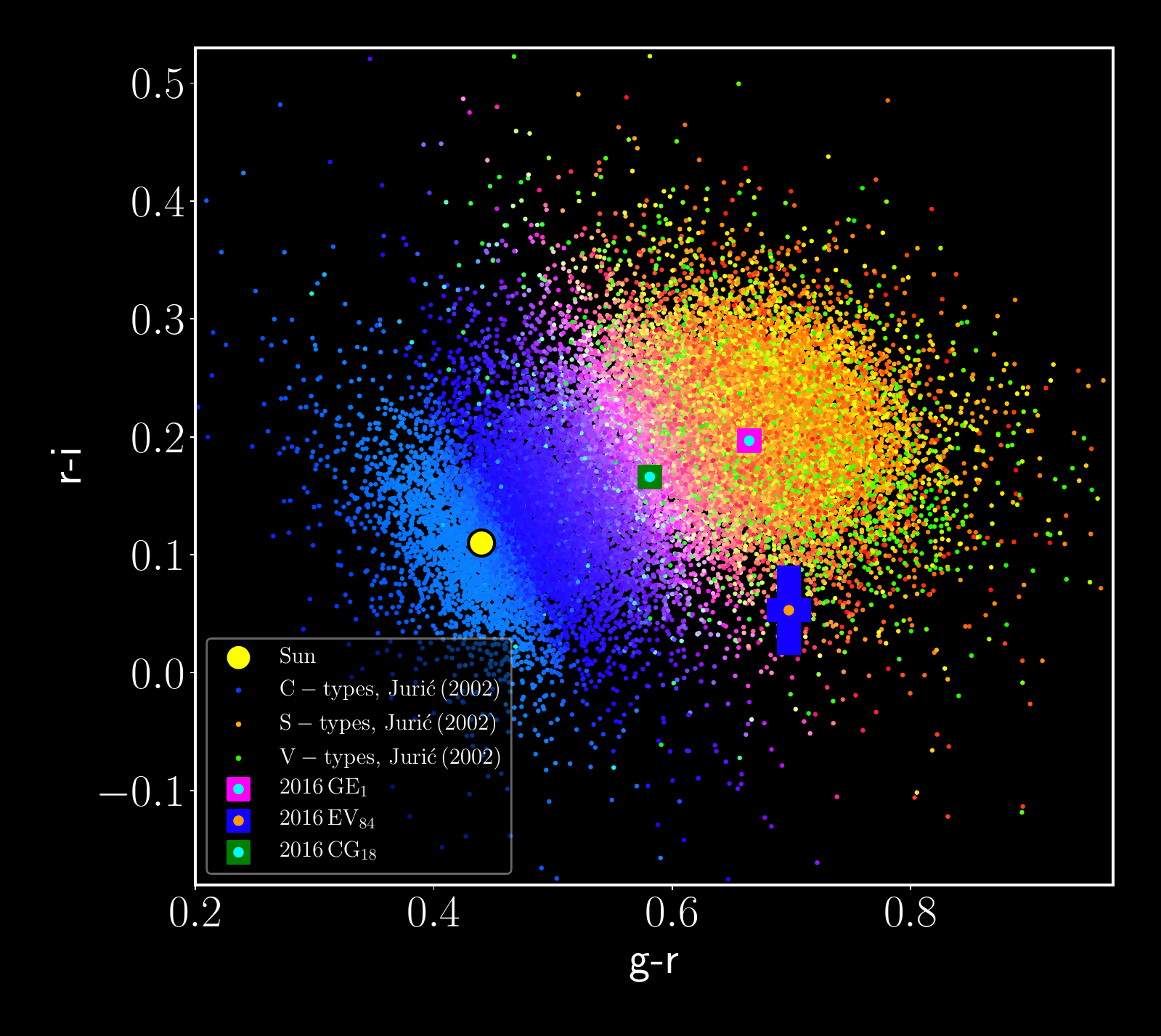}
\caption{g-r vs r-i colours for \geans, \cgans, and \eva superimposed on main belt asteroid colours from the SDSS \citep[][]{Ivezic2001,Juric2002}. The circular yellow marker indicates the Sun's colours (See Table~1.)}
\end{figure}

\begin{table}
\caption{Main belt source region probabilities and albedos.}
\begin{tabular}{llllll}
\hline
object         & $\nu_6$ & 3:1  & Inner & Hungaria & albedo \\ \hline
2016 CG$_{18}$ & 0.75    & 0.14 & 0.02  & 0.09     & 0.21   \\
2016 EV$_{84}$ & 0.72    & 0.17 & 0.05  & 0.06     & 0.20   \\
2016 GE$_1$    & 0.75    & 0.12 & 0.11  & 0.02     & 0.19   \\ \hline
\end{tabular}
\begin{tablenotes}
\item \textbf{Notes.} Main belt source probabilites according the NEO model of \citet{Nesvorny2023NEO} for sources that have $>1$\% probability for these objects. i.e. the $\nu_6$ and 3:1 resonances, the `Inner' main belt source, and the Hungaria family.  Albedos of the objects as predicted by \cite{Morbidelli2020}.
\end{tablenotes}
\end{table}

\begin{figure}
\centering
\includegraphics[width=1\linewidth]{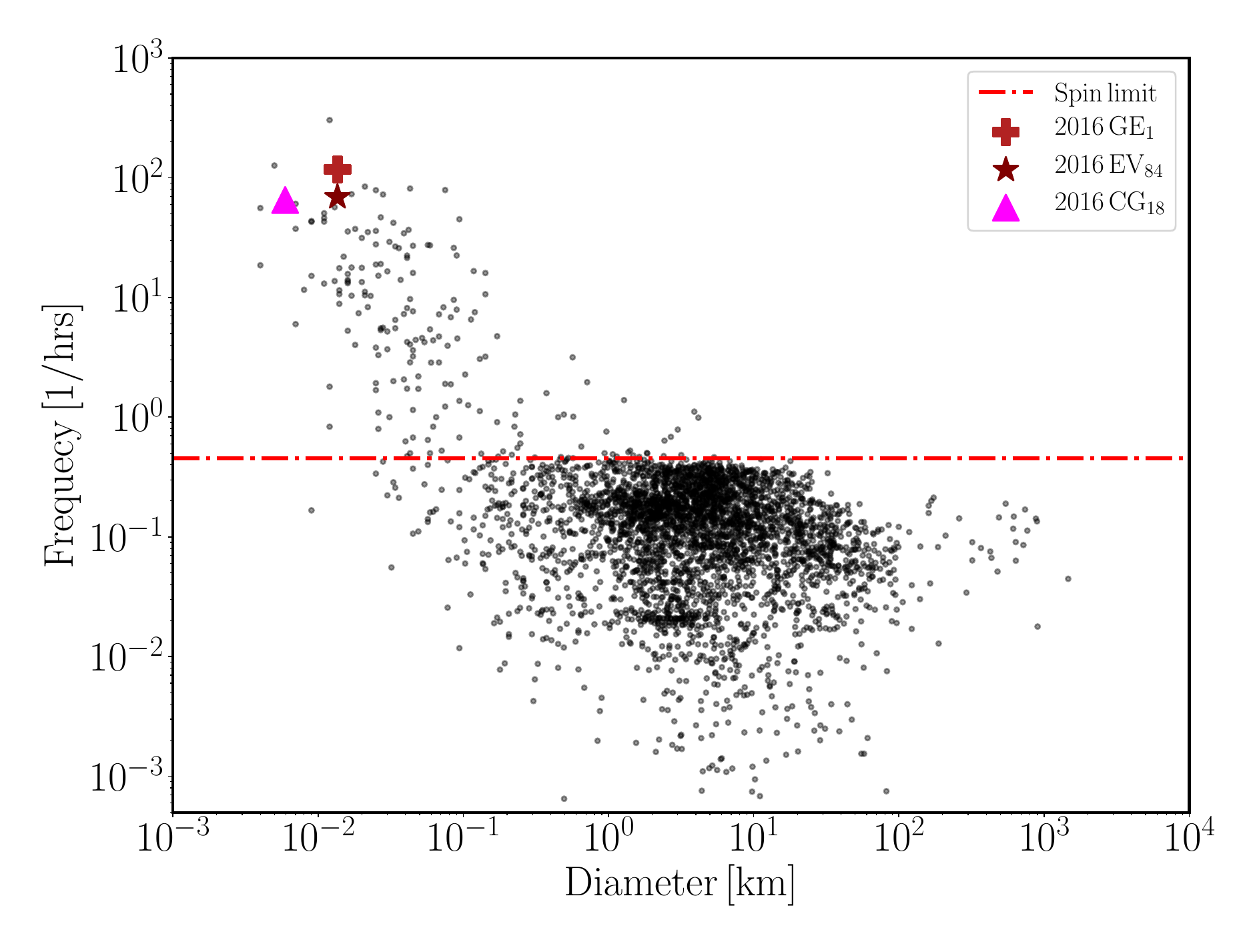}
\caption{Rotation frequency vs. diameter for asteroids in the Asteroid Lightcurve Database with quality codes U=2,3 \citep[][]{Warner2009ab} and the rotation periods for \geans, \cgans, and \evans\ from this work.  The diameters, $D$ of our 3 targets were calculated from their absolute magnitude and an assumed albedo, $p\simeq0.25$, typical of small S-type asteroids \citep{Morbidelli2020,Bolin2023Dink}, with $D = 1329 p^{-1/2} 10^{-H/5}$ \citep{Russell1916}. The spin barrier frequency corresponding to a period of $\sim$2.2~h is indicated with a horizontal red dash-dotted line \citep[][]{Polishook2016}.}
\end{figure}

\section{Conclusions}

We have demonstrated the use of an asteroid streak lightcurve technique on three $\sim$10~m-scale asteroids, \geans, \cgans, and \evans, and compared the results with existing data. The asteroid streak technique may be applicable in the future when using CCD cameras to study fast-rotating asteroids. In addition, the increased lightcurve coverage may improve the measurement of the physical properties of asteroids using optical CCDs that are too difficult to observe by other means due to having a short timespan of observability, e.g. in the case of close-approaching asteroids \citep[][]{Jedicke2018mm,Farnocchia2022}, or proximity to the Sun, e.g. interior-Earth objects \citep[e.g.,][]{Bolin2022IVO,Bolin2023Com}.

\section*{Acknowledgements}
Based on observations obtained with MegaPrime/MegaCam, a joint project of CFHT and CEA/DAPNIA, at the Canada-France-Hawaii Telescope (CFHT) which is operated by the National Research Council (NRC) of Canada, the Institut National des Science de l'Univers of the Centre National de la Recherche Scientifique (CNRS) of France, and the University of Hawaii. The observations at the Canada-France-Hawaii Telescope were performed with care and respect from the summit of Maunakea which is a significant cultural and historic site.
B.T.B. is supported by an appointment to the NASA Postdoctoral Program at the NASA Goddard Space Flight Center, administered by Oak Ridge Associated Universities under contract with NASA. M.G. was supported by the Research Experiences for Undergraduates program at the Institute for Astronomy, Univ. of Hawai`i at M\={a}noa.
\section*{Data Availability}
The data underlying this article will be shared on reasonable request to the corresponding author.
\section*{Supplemental Material}
The supplemental material for this manuscript is available online.



\bibliographystyle{mnras}
\bibliography{/Users/bolin/Dropbox/Projects/NEOZTF_NEOs/neobib} 

\begin{thebibliography}{}
\makeatletter
\relax
\def\mn@urlcharsother{\let\do\@makeother \do\$\do\&\do\#\do\^\do\_\do\%\do\~}
\def\mn@doi{\begingroup\mn@urlcharsother \@ifnextchar [ {\mn@doi@}
  {\mn@doi@[]}}
\def\mn@doi@[#1]#2{\def\@tempa{#1}\ifx\@tempa\@empty \href
  {http://dx.doi.org/#2} {doi:#2}\else \href {http://dx.doi.org/#2} {#1}\fi
  \endgroup}
\def\mn@eprint#1#2{\mn@eprint@#1:#2::\@nil}
\def\mn@eprint@arXiv#1{\href {http://arxiv.org/abs/#1} {{\tt arXiv:#1}}}
\def\mn@eprint@dblp#1{\href {http://dblp.uni-trier.de/rec/bibtex/#1.xml}
  {dblp:#1}}
\def\mn@eprint@#1:#2:#3:#4\@nil{\def\@tempa {#1}\def\@tempb {#2}\def\@tempc
  {#3}\ifx \@tempc \@empty \let \@tempc \@tempb \let \@tempb \@tempa \fi \ifx
  \@tempb \@empty \def\@tempb {arXiv}\fi \@ifundefined
  {mn@eprint@\@tempb}{\@tempb:\@tempc}{\expandafter \expandafter \csname
  mn@eprint@\@tempb\endcsname \expandafter{\@tempc}}}

\bibitem[\protect\citeauthoryear{{Barucci} \& {Fulchignoni}}{{Barucci} \&
  {Fulchignoni}}{1982}]{Barucci1982}
{Barucci} M.~A.,  {Fulchignoni} M.,  1982, Moon and Planets, \href
  {http://adsabs.harvard.edu/abs/1982M%26P....27...47B} {27, 47}

\bibitem[\protect\citeauthoryear{{Beniyama} et~al.,}{{Beniyama}
  et~al.}{2022}]{Beniyama2022}
{Beniyama} J.,  et~al., 2022, \mn@doi [\pasj] {10.1093/pasj/psac043}, \href
  {https://ui.adsabs.harvard.edu/abs/2022PASJ...74..877B} {74, 877}

\bibitem[\protect\citeauthoryear{{Bianco}, {Protopapas}, {McLeod}, {Alcock},
  {Holman}  \& {Lehner}}{{Bianco} et~al.}{2009}]{Bianco2009}
{Bianco} F.~B.,  {Protopapas} P.,  {McLeod} B.~A.,  {Alcock} C.~R.,  {Holman}
  M.~J.,   {Lehner} M.~J.,  2009, \mn@doi [\aj] {10.1088/0004-6256/138/2/568},
  \href {https://ui.adsabs.harvard.edu/abs/2009AJ....138..568B} {138, 568}

\bibitem[\protect\citeauthoryear{{Binzel}, {Farinella}, {Zappal\`a}  \&
  {Cellino}}{{Binzel} et~al.}{1989}]{Binzel1989}
{Binzel} R.~P.,  {Farinella} P.,  {Zappal\`a} V.,   {Cellino} A.,  1989, in
  {Binzel} R.~P.,  {Gehrels} T.,   {Matthews} M.~S.,  eds, Asteroids II. pp
  416--441

\bibitem[\protect\citeauthoryear{{Bolin} \& {Lisse}}{{Bolin} \&
  {Lisse}}{2020}]{Bolin2020HST}
{Bolin} B.~T.,  {Lisse} C.~M.,  2020, \mn@doi [\mnras]
  {10.1093/mnras/staa2192}, \href
  {https://ui.adsabs.harvard.edu/abs/2020MNRAS.497.4031B} {497, 4031}

\bibitem[\protect\citeauthoryear{{Bolin} et~al.,}{{Bolin}
  et~al.}{2014}]{Bolin2014}
{Bolin} B.,  et~al., 2014, \mn@doi [\icarus] {10.1016/j.icarus.2014.05.026},
  \href {http://adsabs.harvard.edu/abs/2014Icar..241..280B} {241, 280}

\bibitem[\protect\citeauthoryear{{Bolin} et~al.,}{{Bolin}
  et~al.}{2018}]{Bolin2018}
{Bolin} B.~T.,  et~al., 2018, \mn@doi [\apjl] {10.3847/2041-8213/aaa0c9}, \href
  {http://adsabs.harvard.edu/abs/2018ApJ...852L...2B} {852, L2}

\bibitem[\protect\citeauthoryear{{Bolin} et~al.,}{{Bolin}
  et~al.}{2020}]{Bolin2020CD3}
{Bolin} B.~T.,  et~al., 2020, \mn@doi [\apjl] {10.3847/2041-8213/abae69}, \href
  {https://ui.adsabs.harvard.edu/abs/2020ApJ...900L..45B} {900, L45}

\bibitem[\protect\citeauthoryear{{Bolin} et~al.,}{{Bolin}
  et~al.}{2022}]{Bolin2022IVO}
{Bolin} B.~T.,  et~al., 2022, \mn@doi [\mnras] {10.1093/mnrasl/slac089}, \href
  {https://ui.adsabs.harvard.edu/abs/2022MNRAS.517L..49B} {517, L49}

\bibitem[\protect\citeauthoryear{{Bolin}, {Ghosal}  \& {Jedicke}}{{Bolin}
  et~al.}{2023a}]{Bolin2023SALsup}
{Bolin} B.~T.,  {Ghosal} M.,   {Jedicke} R.,  2023a, Supplemental Material.

\bibitem[\protect\citeauthoryear{{Bolin}, {Ahumada}, {Dokkum}, {Fremling},
  {Hardegree-Ullman}, {Purdum}, {Serabyn}  \& {Southworth}}{{Bolin}
  et~al.}{2023b}]{Bolin2023Com}
{Bolin} B.~T.,  {Ahumada} T.,  {Dokkum} P.~v.,  {Fremling} C.,
  {Hardegree-Ullman} K.~K.,  {Purdum} J.~N.,  {Serabyn} E.,   {Southworth} J.,
  2023b, \mn@doi [\icarus] {10.1016/j.icarus.2023.115442}, \href
  {https://ui.adsabs.harvard.edu/abs/2023Icar..39415442B} {394, 115442}

\bibitem[\protect\citeauthoryear{{Bolin}, {Noll}, {Caiazzo}, {Fremling}  \&
  {Binzel}}{{Bolin} et~al.}{2023c}]{Bolin2023Dink}
{Bolin} B.~T.,  {Noll} K.~S.,  {Caiazzo} I.,  {Fremling} C.,   {Binzel} R.~P.,
  2023c, \mn@doi [\icarus] {10.1016/j.icarus.2023.115562}, \href
  {https://ui.adsabs.harvard.edu/abs/2023Icar..40015562B} {400, 115562}

\bibitem[\protect\citeauthoryear{{Bolin} et~al.,}{{Bolin}
  et~al.}{2023d}]{Bolin2023NT}
{Bolin} B.~T.,  et~al., 2023d, \mn@doi [\mnras] {10.1093/mnrasl/slad018}, \href
  {https://ui.adsabs.harvard.edu/abs/2023MNRAS.521L..29B} {521, L29}

\bibitem[\protect\citeauthoryear{{Bottke}, {Durda}, {Nesvorn{\'y}}, {Jedicke},
  {Morbidelli}, {Vokrouhlick{\'y}}  \& {Levison}}{{Bottke}
  et~al.}{2005a}]{Bottke2005-fossilized}
{Bottke} W.~F.,  {Durda} D.~D.,  {Nesvorn{\'y}} D.,  {Jedicke} R.,
  {Morbidelli} A.,  {Vokrouhlick{\'y}} D.,   {Levison} H.,  2005a, \mn@doi
  [\icarus] {10.1016/j.icarus.2004.10.026}, \href
  {https://ui.adsabs.harvard.edu/abs/2005Icar..175..111B} {175, 111}

\bibitem[\protect\citeauthoryear{{Bottke}, {Durda}, {Nesvorn{\'y}}, {Jedicke},
  {Morbidelli}, {Vokrouhlick{\'y}}  \& {Levison}}{{Bottke}
  et~al.}{2005b}]{Bottke2005-linking}
{Bottke} W.~F.,  {Durda} D.~D.,  {Nesvorn{\'y}} D.,  {Jedicke} R.,
  {Morbidelli} A.,  {Vokrouhlick{\'y}} D.,   {Levison} H.~F.,  2005b, \mn@doi
  [\icarus] {10.1016/j.icarus.2005.05.017}, \href
  {https://ui.adsabs.harvard.edu/abs/2005Icar..179...63B} {179, 63}

\bibitem[\protect\citeauthoryear{{Bottke}, {Vokrouhlick{\'y}}, {Rubincam}  \&
  {Nesvorn{\'y}}}{{Bottke} et~al.}{2006}]{Bottke2006}
{Bottke} Jr. W.~F.,  {Vokrouhlick{\'y}} D.,  {Rubincam} D.~P.,   {Nesvorn{\'y}}
  D.,  2006, \mn@doi [Annual Review of Earth and Planetary Sciences]
  {10.1146/annurev.earth.34.031405.125154}, \href
  {http://adsabs.harvard.edu/abs/2006AREPS..34..157B} {34, 157}

\bibitem[\protect\citeauthoryear{{Clark}, {Wiegert}, {Brown}, {Vida}, {Heinze}
  \& {Denneau}}{{Clark} et~al.}{2023}]{Clark2023bolide}
{Clark} D.~L.,  {Wiegert} P.~A.,  {Brown} P.~G.,  {Vida} D.,  {Heinze} A.,
  {Denneau} L.,  2023, \mn@doi [\psj] {10.3847/PSJ/acc9b1}, \href
  {https://ui.adsabs.harvard.edu/abs/2023PSJ.....4..103C} {4, 103}

\bibitem[\protect\citeauthoryear{{Cuillandre}, {Magnier}, {Isani}, {Sabin},
  {Knight}, {Kras}  \& {Lai}}{{Cuillandre} et~al.}{2002}]{Cuillandre2002}
{Cuillandre} J.-C.,  {Magnier} E.~A.,  {Isani} S.,  {Sabin} D.,  {Knight} W.,
  {Kras} S.,   {Lai} K.,  2002, in {Quinn} P.~J.,  ed.,  Society of
  Photo-Optical Instrumentation Engineers (SPIE) Conference Series Vol. 4844,
  Observatory Operations to Optimize Scientific Return III. pp 501--507,
  \mn@doi{10.1117/12.460613}

\bibitem[\protect\citeauthoryear{{Daniels}, {Bianco}, {Andreoni}  \&
  {Mahabal}}{{Daniels} et~al.}{2023}]{Daniels2023}
{Daniels} S.,  {Bianco} F.~B.,  {Andreoni} I.,   {Mahabal} A.,  2023, in prep.

\bibitem[\protect\citeauthoryear{{DeMeo} \& {Carry}}{{DeMeo} \&
  {Carry}}{2013}]{DeMeo2013aa}
{DeMeo} F.~E.,  {Carry} B.,  2013, \mn@doi [\icarus]
  {10.1016/j.icarus.2013.06.027}, \href
  {https://ui.adsabs.harvard.edu/abs/2013Icar..226..723D} {226, 723}

\bibitem[\protect\citeauthoryear{{Devog{\`e}le} et~al.,}{{Devog{\`e}le}
  et~al.}{2019}]{Devogele2019}
{Devog{\`e}le} M.,  et~al., 2019, \mn@doi [\aj] {10.3847/1538-3881/ab43dd},
  \href {https://ui.adsabs.harvard.edu/abs/2019AJ....158..196D} {158, 196}

\bibitem[\protect\citeauthoryear{{Farinella}, {Vokrouhlick\'y}  \&
  {Hartmann}}{{Farinella} et~al.}{1998}]{Farinella1998}
{Farinella} P.,  {Vokrouhlick\'y} D.,   {Hartmann} W.~K.,  1998, \icarus, 132,
  378

\bibitem[\protect\citeauthoryear{{Farnocchia} et~al.,}{{Farnocchia}
  et~al.}{2022}]{Farnocchia2022}
{Farnocchia} D.,  et~al., 2022, \mn@doi [\psj] {10.3847/PSJ/ac7224}, \href
  {https://ui.adsabs.harvard.edu/abs/2022PSJ.....3..156F} {3, 156}

\bibitem[\protect\citeauthoryear{{Fraser}, {Pravec}, {Fitzsimmons}, {Lacerda},
  {Bannister}, {Snodgrass}  \& {Smoli{\'c}}}{{Fraser}
  et~al.}{2018}]{Fraser2018}
{Fraser} W.~C.,  {Pravec} P.,  {Fitzsimmons} A.,  {Lacerda} P.,  {Bannister}
  M.~T.,  {Snodgrass} C.,   {Smoli{\'c}} I.,  2018, \mn@doi [Nature Astronomy]
  {10.1038/s41550-018-0398-z}, \href
  {https://ui.adsabs.harvard.edu/abs/2018NatAs...2..383F} {2, 383}

\bibitem[\protect\citeauthoryear{{Fukugita}, {Ichikawa}, {Gunn}, {Doi},
  {Shimasaku}  \& {Schneider}}{{Fukugita} et~al.}{1996}]{Fukugita1996}
{Fukugita} M.,  {Ichikawa} T.,  {Gunn} J.~E.,  {Doi} M.,  {Shimasaku} K.,
  {Schneider} D.~P.,  1996, \mn@doi [\aj] {10.1086/117915}, \href
  {https://ui.adsabs.harvard.edu/abs/1996AJ....111.1748F} {111, 1748}

\bibitem[\protect\citeauthoryear{{Gwyn}, {Hill}  \& {Kavelaars}}{{Gwyn}
  et~al.}{2012}]{Gwyn2012}
{Gwyn} S. D.~J.,  {Hill} N.,   {Kavelaars} J.~J.,  2012, \mn@doi [\pasp]
  {10.1086/666462}, \href
  {https://ui.adsabs.harvard.edu/abs/2012PASP..124..579G} {124, 579}

\bibitem[\protect\citeauthoryear{{Haberreiter}, {Sch{\"o}ll}, {Dudok de Wit},
  {Kretzschmar}, {Misios}, {Tourpali}  \& {Schmutz}}{{Haberreiter}
  et~al.}{2017}]{Haberreiter2017}
{Haberreiter} M.,  {Sch{\"o}ll} M.,  {Dudok de Wit} T.,  {Kretzschmar} M.,
  {Misios} S.,  {Tourpali} K.,   {Schmutz} W.,  2017, \mn@doi [Journal of
  Geophysical Research (Space Physics)] {10.1002/2016JA023492}, \href
  {https://ui.adsabs.harvard.edu/abs/2017JGRA..122.5910H} {122, 5910}

\bibitem[\protect\citeauthoryear{{Hanu{\v s}} et~al.,}{{Hanu{\v s}}
  et~al.}{2018}]{Hanus2018}
{Hanu{\v s}} J.,  et~al., 2018, \mn@doi [\icarus]
  {10.1016/j.icarus.2017.07.007}, \href
  {http://adsabs.harvard.edu/abs/2018Icar..299...84H} {299, 84}

\bibitem[\protect\citeauthoryear{{Harding} et~al.,}{{Harding}
  et~al.}{2016}]{Harding2016}
{Harding} L.~K.,  et~al., 2016, \mn@doi [\mnras] {10.1093/mnras/stw094}, \href
  {https://ui.adsabs.harvard.edu/abs/2016MNRAS.457.3036H} {457, 3036}

\bibitem[\protect\citeauthoryear{{Harris} \& {Lagerros}}{{Harris} \&
  {Lagerros}}{2002}]{Harris2002}
{Harris} A.~W.,  {Lagerros} J.~S.~V.,  2002, Asteroids III, \href
  {http://adsabs.harvard.edu/abs/2002aste.book..205H} {pp 205--218}

\bibitem[\protect\citeauthoryear{{Harris}, {Fahnestock}  \& {Pravec}}{{Harris}
  et~al.}{2009}]{Harris2009a}
{Harris} A.~W.,  {Fahnestock} E.~G.,   {Pravec} P.,  2009, \mn@doi [\icarus]
  {10.1016/j.icarus.2008.09.012}, \href
  {http://adsabs.harvard.edu/abs/2009Icar..199..310H} {199, 310}

\bibitem[\protect\citeauthoryear{{Hartman}, {Bersier}, {Stanek}, {Beaulieu},
  {Kaluzny}, {Marquette}, {Stetson}  \& {Schwarzenberg-Czerny}}{{Hartman}
  et~al.}{2006}]{Hartman2006}
{Hartman} J.~D.,  {Bersier} D.,  {Stanek} K.~Z.,  {Beaulieu} J.~P.,  {Kaluzny}
  J.,  {Marquette} J.~B.,  {Stetson} P.~B.,   {Schwarzenberg-Czerny} A.,  2006,
  \mn@doi [\mnras] {10.1111/j.1365-2966.2006.10764.x}, \href
  {https://ui.adsabs.harvard.edu/abs/2006MNRAS.371.1405H} {371, 1405}

\bibitem[\protect\citeauthoryear{{Hirabayashi}}{{Hirabayashi}}{2015}]{Hirabayashi2015}
{Hirabayashi} M.,  2015, \mn@doi [\mnras] {10.1093/mnras/stv2017}, \href
  {http://adsabs.harvard.edu/abs/2015MNRAS.454.2249H} {454, 2249}

\bibitem[\protect\citeauthoryear{{Ivezi{\'c}} et~al.,}{{Ivezi{\'c}}
  et~al.}{2001}]{Ivezic2001}
{Ivezi{\'c}} {\v Z}.,  et~al., 2001, \mn@doi [\aj] {10.1086/323452}, \href
  {http://adsabs.harvard.edu/abs/2001AJ....122.2749I} {122, 2749}

\bibitem[\protect\citeauthoryear{{Jedicke}, {Granvik}, {Micheli}, {Ryan},
  {Spahr}  \& {Yeomans}}{{Jedicke} et~al.}{2015}]{Jedicke2015}
{Jedicke} R.,  {Granvik} M.,  {Micheli} M.,  {Ryan} E.,  {Spahr} T.,
  {Yeomans} D.~K.,  2015, \mn@doi [Asteroids IV]
  {10.2458/azu_uapress_9780816532131-ch040}, \href
  {http://adsabs.harvard.edu/abs/2015aste.book..795J} {pp 795--813}

\bibitem[\protect\citeauthoryear{{Jedicke}, {Bolin}, {Granvik}  \&
  {Beshore}}{{Jedicke} et~al.}{2016}]{Jedicke2016}
{Jedicke} R.,  {Bolin} B.,  {Granvik} M.,   {Beshore} E.,  2016, \mn@doi
  [\icarus] {10.1016/j.icarus.2015.10.021}, \href
  {http://adsabs.harvard.edu/abs/2016Icar..266..173J} {266, 173}

\bibitem[\protect\citeauthoryear{{Jedicke}, {Bolin}, {Bottke}, {Chyba},
  {Fedorets}, {Granvik}, {Jones}  \& {Urrutxua}}{{Jedicke}
  et~al.}{2018a}]{Jedicke2018mm}
{Jedicke} R.,  {Bolin} B.~T.,  {Bottke} W.~F.,  {Chyba} M.,  {Fedorets} G.,
  {Granvik} M.,  {Jones} L.,   {Urrutxua} H.,  2018a, \mn@doi [Frontiers in
  Astronomy and Space Sciences] {10.3389/fspas.2018.00013}, \href
  {https://ui.adsabs.harvard.edu/abs/2018FrASS...5...13J} {5, 13}

\bibitem[\protect\citeauthoryear{{Jedicke}, {Sercel}, {Gillis-Davis}, {Morenz}
  \& {Gertsch}}{{Jedicke} et~al.}{2018b}]{Jedicke2018}
{Jedicke} R.,  {Sercel} J.,  {Gillis-Davis} J.,  {Morenz} K.~J.,   {Gertsch}
  L.,  2018b, \mn@doi [\planss] {10.1016/j.pss.2018.04.005}, \href
  {https://ui.adsabs.harvard.edu/abs/2018P&SS..159...28J} {159, 28}

\bibitem[\protect\citeauthoryear{{Jenniskens} et~al.,}{{Jenniskens}
  et~al.}{2010}]{Jenniskens2010}
{Jenniskens} P.,  et~al., 2010, \mn@doi [Meteoritics and Planetary Science]
  {10.1111/j.1945-5100.2010.01153.x}, \href
  {http://adsabs.harvard.edu/abs/2010M%26PS...45.1590J} {45, 1590}

\bibitem[\protect\citeauthoryear{{Juri{\'c}} et~al.,}{{Juri{\'c}}
  et~al.}{2002}]{Juric2002}
{Juri{\'c}} M.,  et~al., 2002, \mn@doi [\aj] {10.1086/341950}, \href
  {http://adsabs.harvard.edu/abs/2002AJ....124.1776J} {124, 1776}

\bibitem[\protect\citeauthoryear{{Kaasalainen}}{{Kaasalainen}}{2001}]{Kaasalainen2001}
{Kaasalainen} M.,  2001, \mn@doi [\aap] {10.1051/0004-6361:20010935}, \href
  {https://ui.adsabs.harvard.edu/abs/2001A&A...376..302K} {376, 302}

\bibitem[\protect\citeauthoryear{{Kozubal}, {Gasdia}, {Dantowitz}, {Scheirich}
  \& {Harris}}{{Kozubal} et~al.}{2011}]{Kozubal2011}
{Kozubal} M.~J.,  {Gasdia} F.~W.,  {Dantowitz} R.~F.,  {Scheirich} P.,
  {Harris} A.~W.,  2011, \mn@doi [\maps] {10.1111/j.1945-5100.2011.01172.x},
  \href {https://ui.adsabs.harvard.edu/abs/2011M&PS...46..534K} {46, 534}

\bibitem[\protect\citeauthoryear{{Lindberg} et~al.,}{{Lindberg}
  et~al.}{2022}]{Lindberg2022}
{Lindberg} C.~W.,  et~al., 2022, \mn@doi [\aj] {10.3847/1538-3881/ac3079},
  \href {https://ui.adsabs.harvard.edu/abs/2022AJ....163...29L} {163, 29}

\bibitem[\protect\citeauthoryear{{Lomb}}{{Lomb}}{1976}]{Lomb1976}
{Lomb} N.~R.,  1976, \mn@doi [\apss] {10.1007/BF00648343}, \href
  {http://adsabs.harvard.edu/abs/1976Ap%26SS..39..447L} {39, 447}

\bibitem[\protect\citeauthoryear{{Magnier} \& {Cuillandre}}{{Magnier} \&
  {Cuillandre}}{2004}]{Magnier2004}
{Magnier} E.~A.,  {Cuillandre} J.~C.,  2004, \mn@doi [\pasp] {10.1086/420756},
  \href {https://ui.adsabs.harvard.edu/abs/2004PASP..116..449M} {116, 449}

\bibitem[\protect\citeauthoryear{{McNeill} et~al.,}{{McNeill}
  et~al.}{2018}]{McNeill2018}
{McNeill} A.,  et~al., 2018, \mn@doi [\aj] {10.3847/1538-3881/aaeb8c}, \href
  {https://ui.adsabs.harvard.edu/abs/2018AJ....156..282M} {156, 282}

\bibitem[\protect\citeauthoryear{{Micheli}, {Tholen}  \& {Elliott}}{{Micheli}
  et~al.}{2013}]{Micheli2013-2012LA}
{Micheli} M.,  {Tholen} D.~J.,   {Elliott} G.~T.,  2013, \mn@doi [\icarus]
  {10.1016/j.icarus.2013.05.032}, \href
  {https://ui.adsabs.harvard.edu/abs/2013Icar..226..251M} {226, 251}

\bibitem[\protect\citeauthoryear{{Micheli}, {Tholen}  \& {Elliott}}{{Micheli}
  et~al.}{2014}]{Micheli2014-2011MD}
{Micheli} M.,  {Tholen} D.~J.,   {Elliott} G.~T.,  2014, \mn@doi [\apjl]
  {10.1088/2041-8205/788/1/L1}, \href
  {https://ui.adsabs.harvard.edu/abs/2014ApJ...788L...1M} {788, L1}

\bibitem[\protect\citeauthoryear{{Morbidelli}, {Delbo}, {Granvik}, {Bottke},
  {Jedicke}, {Bolin}, {Michel}  \& {Vokrouhlicky}}{{Morbidelli}
  et~al.}{2020}]{Morbidelli2020}
{Morbidelli} A.,  {Delbo} M.,  {Granvik} M.,  {Bottke} W.~F.,  {Jedicke} R.,
  {Bolin} B.,  {Michel} P.,   {Vokrouhlicky} D.,  2020, \mn@doi [\icarus]
  {10.1016/j.icarus.2020.113631}, \href
  {https://ui.adsabs.harvard.edu/abs/2020Icar..34013631M} {340, 113631}

\bibitem[\protect\citeauthoryear{{Nesvorn{\'y}} et~al.,}{{Nesvorn{\'y}}
  et~al.}{2023}]{Nesvorny2023NEO}
{Nesvorn{\'y}} D.,  et~al., 2023, \mn@doi [\aj] {10.3847/1538-3881/ace040},
  \href {https://ui.adsabs.harvard.edu/abs/2023AJ....166...55N} {166, 55}

\bibitem[\protect\citeauthoryear{{Polishook} et~al.,}{{Polishook}
  et~al.}{2016}]{Polishook2016}
{Polishook} D.,  et~al., 2016, \mn@doi [\icarus]
  {10.1016/j.icarus.2015.12.031}, \href
  {https://ui.adsabs.harvard.edu/abs/2016Icar..267..243P} {267, 243}

\bibitem[\protect\citeauthoryear{{Pomazan}, {Tang}, {Maigurova}, {Yu}, {Tang},
  {Mao}  \& {Song}}{{Pomazan} et~al.}{2022}]{Pomazan2022}
{Pomazan} A.,  {Tang} Z.-H.,  {Maigurova} N.,  {Yu} Y.,  {Tang} K.,  {Mao}
  Y.-D.,   {Song} Y.-Z.,  2022, \mn@doi [\planss] {10.1016/j.pss.2022.105477},
  \href {https://ui.adsabs.harvard.edu/abs/2022P&SS..21605477P} {216, 105477}

\bibitem[\protect\citeauthoryear{{Pravec} et~al.,}{{Pravec}
  et~al.}{2005}]{Pravec2005}
{Pravec} P.,  et~al., 2005, \mn@doi [\icarus] {10.1016/j.icarus.2004.07.021},
  \href {https://ui.adsabs.harvard.edu/abs/2005Icar..173..108P} {173, 108}

\bibitem[\protect\citeauthoryear{{Press}, {Flannery}  \& {Teukolsky}}{{Press}
  et~al.}{1986}]{Press1986}
{Press} W.~H.,  {Flannery} B.~P.,   {Teukolsky} S.~A.,  1986, {Numerical
  recipes. The art of scientific computing}

\bibitem[\protect\citeauthoryear{{Purdum} et~al.,}{{Purdum}
  et~al.}{2021}]{Purdum2021}
{Purdum} J.~N.,  et~al., 2021, \mn@doi [\apjl] {10.3847/2041-8213/abf2ca},
  \href {https://ui.adsabs.harvard.edu/abs/2021ApJ...911L..35P} {911, L35}

\bibitem[\protect\citeauthoryear{{Russell}}{{Russell}}{1916}]{Russell1916}
{Russell} H.~N.,  1916, \mn@doi [\apj] {10.1086/142244}, \href
  {https://ui.adsabs.harvard.edu/abs/1916ApJ....43..173R} {43, 173}

\bibitem[\protect\citeauthoryear{{Ryan}, {Sharkey}  \& {Woodward}}{{Ryan}
  et~al.}{2017}]{Ryan2017}
{Ryan} E.~L.,  {Sharkey} B. N.~L.,   {Woodward} C.~E.,  2017, \mn@doi [\aj]
  {10.3847/1538-3881/153/3/116}, \href
  {https://ui.adsabs.harvard.edu/abs/2017AJ....153..116R} {153, 116}

\bibitem[\protect\citeauthoryear{{S\'{a}nchez} \& {Scheeres}}{{S\'{a}nchez} \&
  {Scheeres}}{2014}]{Sanchez2014}
{S\'{a}nchez} P.,  {Scheeres} D.~J.,  2014, Meteoritics and Planetary Science,
  49, 788

\bibitem[\protect\citeauthoryear{{Scargle}}{{Scargle}}{1982}]{Scargle1982}
{Scargle} J.~D.,  1982, \mn@doi [\apj] {10.1086/160554}, \href
  {http://adsabs.harvard.edu/abs/1982ApJ...263..835S} {263, 835}

\bibitem[\protect\citeauthoryear{{Shao}, {Nemati}, {Zhai}, {Turyshev},
  {Sandhu}, {Hallinan}  \& {Harding}}{{Shao} et~al.}{2014}]{Shao2014}
{Shao} M.,  {Nemati} B.,  {Zhai} C.,  {Turyshev} S.~G.,  {Sandhu} J.,
  {Hallinan} G.,   {Harding} L.~K.,  2014, \mn@doi [ApJ]
  {10.1088/0004-637X/782/1/1}, \href
  {http://adsabs.harvard.edu/abs/2014ApJ...782....1S} {782, 1}

\bibitem[\protect\citeauthoryear{{Sharma} et~al.,}{{Sharma}
  et~al.}{2023}]{Sharma2023}
{Sharma} K.,  et~al., 2023, \mn@doi [\mnras] {10.1093/mnras/stad1989}, \href
  {https://ui.adsabs.harvard.edu/abs/2023MNRAS.524.2651S} {524, 2651}

\bibitem[\protect\citeauthoryear{{Szab{\'o}} et~al.,}{{Szab{\'o}}
  et~al.}{2016}]{Szabo2016}
{Szab{\'o}} R.,  et~al., 2016, \mn@doi [\aap] {10.1051/0004-6361/201629059},
  \href {https://ui.adsabs.harvard.edu/abs/2016A&A...596A..40S} {596, A40}

\bibitem[\protect\citeauthoryear{{Szab{\'o}} et~al.,}{{Szab{\'o}}
  et~al.}{2017}]{Szabo2017}
{Szab{\'o}} G.~M.,  et~al., 2017, \mn@doi [\aap] {10.1051/0004-6361/201629401},
  \href {https://ui.adsabs.harvard.edu/abs/2017A&A...599A..44S} {599, A44}

\bibitem[\protect\citeauthoryear{{Thirouin} \& {Sheppard}}{{Thirouin} \&
  {Sheppard}}{2019}]{Thirouin2019}
{Thirouin} A.,  {Sheppard} S.~S.,  2019, \mn@doi [\aj]
  {10.3847/1538-3881/ab18a9}, \href
  {https://ui.adsabs.harvard.edu/abs/2019AJ....157..228T} {157, 228}

\bibitem[\protect\citeauthoryear{{Thirouin} \& {Sheppard}}{{Thirouin} \&
  {Sheppard}}{2022}]{Thirouin2022}
{Thirouin} A.,  {Sheppard} S.~S.,  2022, \mn@doi [\psj] {10.3847/PSJ/ac7ab8},
  \href {https://ui.adsabs.harvard.edu/abs/2022PSJ.....3..178T} {3, 178}

\bibitem[\protect\citeauthoryear{{Thirouin} et~al.,}{{Thirouin}
  et~al.}{2016}]{Thirouin2016}
{Thirouin} A.,  et~al., 2016, \mn@doi [\aj] {10.3847/0004-6256/152/6/163},
  \href {http://adsabs.harvard.edu/abs/2016AJ....152..163T} {152, 163}

\bibitem[\protect\citeauthoryear{{Vere\v{s}}, {Jedicke}, {Denneau},
  {Wainscoat}, {Holman}  \& {Lin}}{{Vere\v{s}} et~al.}{2012a}]{Veres2012}
{Vere\v{s}} P.,  {Jedicke} R.,  {Denneau} L.,  {Wainscoat} R.,  {Holman} M.~J.,
    {Lin} H.,  2012a, \pasp

\bibitem[\protect\citeauthoryear{{Vere{\v{s}}}, {Jedicke}, {Denneau},
  {Wainscoat}, {Holman}  \& {Lin}}{{Vere{\v{s}}} et~al.}{2012b}]{Veres2012ab}
{Vere{\v{s}}} P.,  {Jedicke} R.,  {Denneau} L.,  {Wainscoat} R.,  {Holman}
  M.~J.,   {Lin} H.-W.,  2012b, \mn@doi [\pasp] {10.1086/668616}, \href
  {https://ui.adsabs.harvard.edu/abs/2012PASP..124.1197V} {124, 1197}

\bibitem[\protect\citeauthoryear{Virtanen et~al.,}{Virtanen
  et~al.}{2020}]{Scipy2020}
Virtanen P.,  et~al., 2020, \mn@doi [Nature Methods]
  {10.1038/s41592-019-0686-2}, \href {https://rdcu.be/b08Wh} {17, 261}

\bibitem[\protect\citeauthoryear{{Vokrouhlick{\'y}} \& {{\v
  C}apek}}{{Vokrouhlick{\'y}} \& {{\v C}apek}}{2002}]{Vokrouhlicky2002}
{Vokrouhlick{\'y}} D.,  {{\v C}apek} D.,  2002, \mn@doi [Icarus]
  {10.1006/icar.2002.6918}, \href
  {http://adsabs.harvard.edu/abs/2002Icar..159..449V} {159, 449}

\bibitem[\protect\citeauthoryear{{Vokrouhlick{\'y}}, {Bottke}, {Chesley},
  {Scheeres}  \& {Statler}}{{Vokrouhlick{\'y}} et~al.}{2015}]{Vokrouhlicky2015}
{Vokrouhlick{\'y}} D.,  {Bottke} W.~F.,  {Chesley} S.~R.,  {Scheeres} D.~J.,
  {Statler} T.~S.,  2015, \mn@doi [Asteroids IV]
  {10.2458/azu_uapress_9780816530595-ch027}, \href
  {http://adsabs.harvard.edu/abs/2015aste.book..509V} {pp 509--531}

\bibitem[\protect\citeauthoryear{{Warner}}{{Warner}}{2016}]{Warner2016GE1}
{Warner} B.~D.,  2016, Minor Planet Bulletin, \href
  {https://ui.adsabs.harvard.edu/abs/2016MPBu...43..240W} {43, 240}

\bibitem[\protect\citeauthoryear{{Warner}, {Harris}  \& {Pravec}}{{Warner}
  et~al.}{2009}]{Warner2009ab}
{Warner} B.~D.,  {Harris} A.~W.,   {Pravec} P.,  2009, \mn@doi [\icarus]
  {10.1016/j.icarus.2009.02.003}, \href
  {https://ui.adsabs.harvard.edu/abs/2009Icar..202..134W} {202, 134}

\bibitem[\protect\citeauthoryear{{Whidden} et~al.,}{{Whidden}
  et~al.}{2019}]{Whidden2019}
{Whidden} P.~J.,  et~al., 2019, \mn@doi [\aj] {10.3847/1538-3881/aafd2d}, \href
  {https://ui.adsabs.harvard.edu/abs/2019AJ....157..119W} {157, 119}

\bibitem[\protect\citeauthoryear{{Willmer}}{{Willmer}}{2018}]{Willmer2018}
{Willmer} C. N.~A.,  2018, \mn@doi [\apjs] {10.3847/1538-4365/aabfdf}, \href
  {https://ui.adsabs.harvard.edu/abs/2018ApJS..236...47W} {236, 47}

\makeatother
\end{thebibliography}




\renewcommand{\thefigure}{S\arabic{figure}}
\setcounter{figure}{0}
\renewcommand{\thetable}{S\arabic{table}}
\renewcommand{\theequation}{S\arabic{equation}}
\renewcommand{\thesection}{S\arabic{section}}
\setcounter{section}{0}
\cleardoublepage
\setcounter{page}{1}
\renewcommand\thepage{S\arabic{page}}

\section*{Supplemental Material}

\appendix
\renewcommand{\thefigure}{S\arabic{figure}}
\setcounter{figure}{0}
\renewcommand{\thetable}{S\arabic{table}}
\renewcommand{\theequation}{S\arabic{equation}}
\renewcommand{\thesection}{S}
\setcounter{section}{0}

In this Supplemental Material section, we provide a detailed mathematical description of the methodology of our technique for extracting and calibrating asteroid streak photometry (Section~S1), the application of the method to real asteroid streak data (Section~S2), and the determination of asteroid rotation period and colours (Section~S3). The methodological description is presented as a general guideline leaving the exact implementation of the streak extraction method to the needs of the user. The asteroid lightcurves in Section~S1 are presented to illustrate the periodic variations in their brightness. The periodogram and colour analysis described in Section~S3 details the application of our method and results for each of our asteroid targets.

\subsection{Asteroid trail extraction and calibration}

Each image was visually inspected to identify the asteroid trail and measure the approximate coordinates of its centroid. We then fit the asteroid source to a Gaussian point spread function convolved with a straight line following \citet{Veres2012}:
\begin{equation}
\label{trailfit}
\begin{split}
f (x, y) =  &\:B + 
\frac{\Phi}{ 2\sigma L  \sqrt{2\pi} }\\
&\exp(- \frac{ \big( (x - x_0 ) \sin{\phi_0} + (y - y_0) \cos{\phi_0} \big)^2 }{2 \sigma^2} )         \\
&\big[ \mathrm{erf}(  \frac{  (x - x_0 ) \cos{\phi_0} + (y - y_0) \sin{\phi_0} + L/2 }{\sigma\sqrt2} ) -  \\
&\mathrm{erf}(  \frac{  (x - x_0 ) \cos{\phi_0} + (y - y_0) \sin{\phi_0} - L/2 }{\sigma\sqrt2}) \big]
\end{split}
\end{equation}
where $(x_0, y_0)$ represent the pixel coordinates of the trail's centroid, $\phi_0$ is the angle made by the trail relative to the positive $x$-axis, $L$ is the trail length, $\sigma$ is the standard deviation of the Gaussian representing the point spread function (PSF),  $\Phi$ is the total flux in the trail, and $B$ represents the background sky.
The fit was obtained with {\tt Scipy.optimize.curve\_fit} which also provides the 1-$\sigma$ uncertainties on the fitted parameters.

As an example, the center of \gea's trail varies from row to row as a consequence of the imperfect tracking of the telescope and time-varying and localized seeing (top panel of Fig.~S1) but {\tt curve\_fit} returns the row-averaged centroid (middle panel of Fig.~S1). The fitted trail defines the trail start and ending points and the trail width necessary for reducing the light curve.

\begin{figure}
\centering 
\includegraphics[width=1\linewidth]{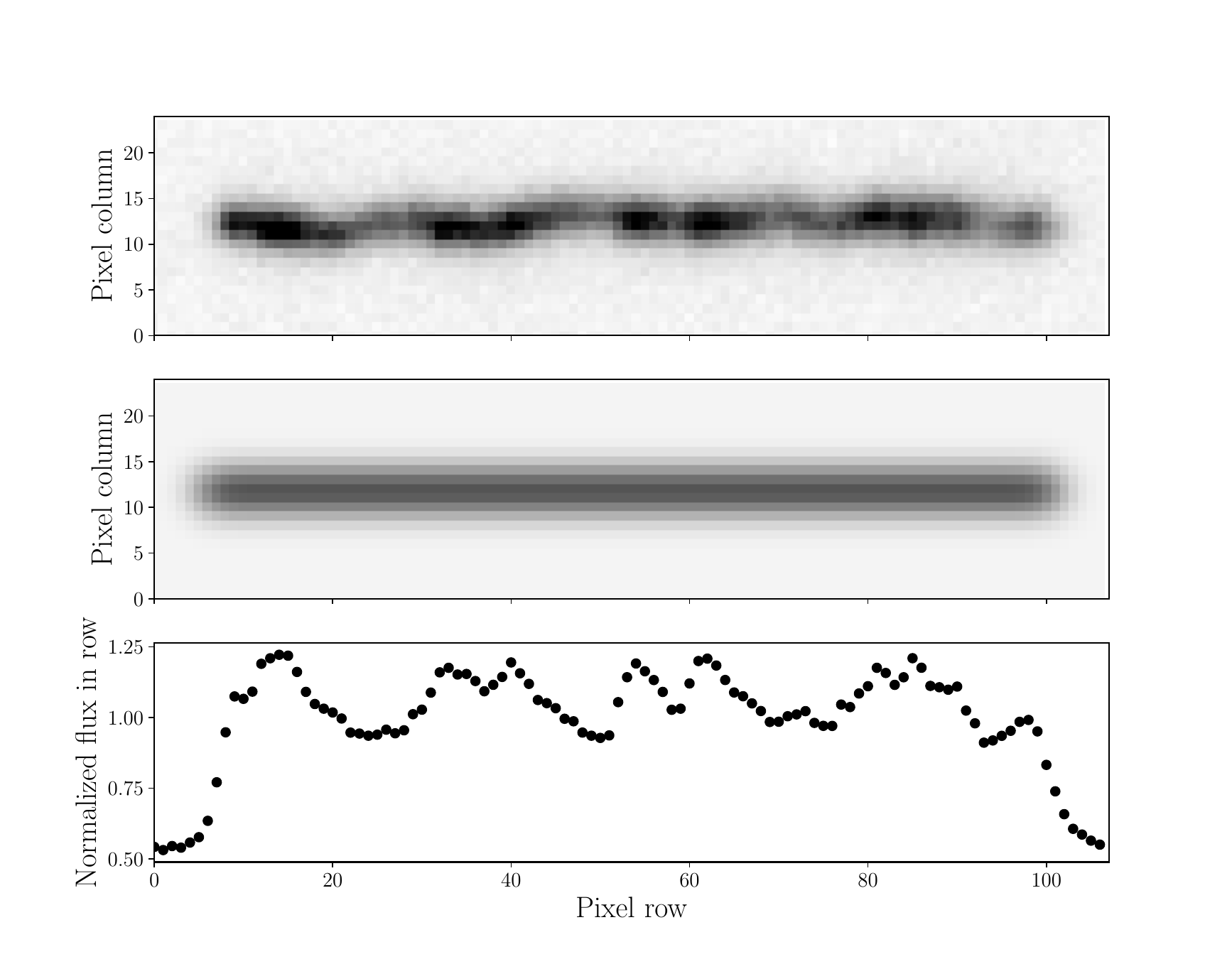}
 \caption{Top panel: asteroid \gea trail from 4 April 2016 rotated by $90^{\circ}$. Middle panel: the same trail as realized from the \citet{Veres2012} trail fit Eq.~\ref{trailfit}. The wiggling appearance of the trail in the perpendicular direction is due to the imperfect tracking of the telescope. Bottom panel: the sum of the flux in each column versus column number normalized by the median of the lightcurve.  The flux has not yet been corrected for sky background and transparency.  The ends of the lightcurve fall to $\sim0.5$ because they sample the background sky.}
\end{figure}

Prior to extracting lightcurves from the asteroid trails, each frame was rotated by $-\phi_0$ degrees using the flux-conserving routine provided by {\tt Scipy.ndimage.rotate} \citep[][]{Scipy2020} resulting in asteroid trails exactly aligned with the CCD columns. $\phi_0$ was not $0^{\circ}$ or $90^{\circ}$ because the orbits of the NEOs at the time of observation were not well constrained so our calculation of the non-sidereal tracking rates could introduce a small tracking error resulting in a misaligned trail.

Once the trails were oriented along CCD columns the row number is linearly correlated with time. The physical rotation of the asteroid will result in a time-varying flux along the rows if the asteroid is rotating and not spherical or has surface albedo variations. Non-sidereal tracking errors could also appear as a small time-varying flux, though inspection of nearby star trails shows this to be smaller than a few percent.

While the fitted trail parameters are real numbers, our light curve reduction is integer pixel-based. 
We define the integer number of pixels in one full-width-at-half-maximum (FWHM) of the PSF as  $n_w = \mathrm{ceil}(2.355 \; \sigma)$.
The fitted trail's endpoints are given by 
\begin{eqnarray}
    (x_1, y_1) &=& (x_0, y_0 + L/2) \\
    (x_2, y_2) &=& (x_0, y_0 - L/2)
\end{eqnarray}
but we truncate the trail on each end by $n_w$ pixels to eliminate edge effects at the beginning and end of the exposure.  
Thus, the effective trail length is $L = L_0 - 2 n_w$ pixels.

Letting $i_n =\mathrm{int}(x_n)$ represent the column/pixel number containing the position $x_n$ and $j_n =\mathrm{int}(y_n)$ represent the row number containing the position $y_n$, the integer pixel bounds for the asteroid's signal are: 
\begin{equation}
    \label{eqn-signal-bounds}
    \begin{split}
        i_{min} &= \mathrm{floor}( \; x_0 - n_w \; )  \\
        i_{max} &= \mathrm{ceil}(  \; x_0 + n_w \; )  \\
        j_{min} &= \mathrm{floor}( \; y_0 - L/2 \; )  \\
        j_{max} &= \mathrm{ceil}(  \; y_0 + L/2 \; )  \\
    \end{split}
\end{equation}
The total flux in each row $m$ through $n$ inclusive is:
\begin{equation}
    \label{eqn-signal}
    \begin{split}
        F_{j} = \sum_{j=m}^{n} \sum_{i=i_{min} }^{ i_{max}} F_{ij}
    \end{split}
\end{equation}
where $F_{ij}$ is the flux in pixel $(i,j)$.

The contribution of the sky background to the asteroid flux is calculated using a set of pixels adjacent to the streak. 
The bounds in $i$ are identical to those in Eq.~\ref{eqn-signal-bounds} but the bounds in $j$ are:
\begin{equation}
    \label{eqn-sky-bounds}
    \begin{split}
        i_{-min} &= i_0 - n_w  \\
        i_{-max} &= i_0 - n_{sky  }  \\
        i_{+min} &= i_0 + n_w  \\
        i_{+max} &= i_0 + n_{sky  }  \\
    \end{split}
\end{equation}
where $n_{sky} = 4 \; n_w$ and the $-$ and $+$ subscripts denote the sky region left and right of the trail, respectively. 
The total number of sky pixels is then $N_{sky} = 2 (n_{sky} - n_w)$ and the total flux in the sky background region in row $j$ bracketing the asteroid trail is:
\begin{equation}
    \label{eqn-signal-sky}
    \begin{split}
        F_{sky, j} = \sum_{i=i_{-min} }^{ i_{-max}} F_{ij} 
                   + \sum_{i=i_{+min} }^{ i_{+max}} F_{ij}.
    \end{split}
\end{equation}
The average background flux per pixel in the region surrounding, and presumably beneath, the trail is then $\Bar{F}_{sky, j} = F_{sky, j}/N_{sky}$. 
Finally, the background-corrected signal in the $j^{th}$ row of the trail is: 
\begin{equation}
    \label{eqn-sobj-minus-sky}
       S_j = F_j - N_{sky}  \; \Bar{F}_{sky, j}
\end{equation}

The error on the background-subtracted flux as a function of row number is \begin{equation}
    \label{eqn-error-flux}
    \begin{split}
        \sigma_{j}^2 = \frac{F_j}{G} + n_w  (\frac{\Bar{F}_{sky, j}}{G} + R^2) + n_w^2  \sigma_{sky,j}^2
    \end{split}
\end{equation}
where $G$ is the CCD gain, the number of CCD `counts' per $e^-$,  $R$ is the read noise of the electronics in $e^-$/pixel, and $\sigma_{sky,j} = \sqrt{F_{sky,j}}/N_{sky}$.
The calibrated magnitude in the $j^{th}$ row of the trail is:
\begin{equation}
    \label{eqn-magnitude}
       M_j = -2.5 * log_{10} ( S_j ) + ZP
\end{equation}
where $ZP$ is the zeropoint obtained from Elixir photometry. 

\subsection{Asteroid streak lightcurves}

The asteroid streak lightcurve extraction procedure described in Section S1 was applied to the g, r, and i images taken for \geans, \cgans, and \eva (observational circumstances described in Table S1). 

The combined g, r and i lightcurves for \gea (top panel of Fig.~S2) show clear evidence of the asteroid's colours (discussed below) while the individual g, r and i lightcurves (panels 2-4 of Fig.~S2) show the sinusoidal shape typical for a prolate spinning body \citep[][]{Barucci1982}. The amplitude in the g and r lightcurve data is $\sim$0.8 magnitudes though it is reduced in the i band, possibly due to tumbling of the asteroid \citep[e.g.,][]{Fraser2018}.

\begin{figure}
\centering 
\includegraphics[width=1\linewidth]{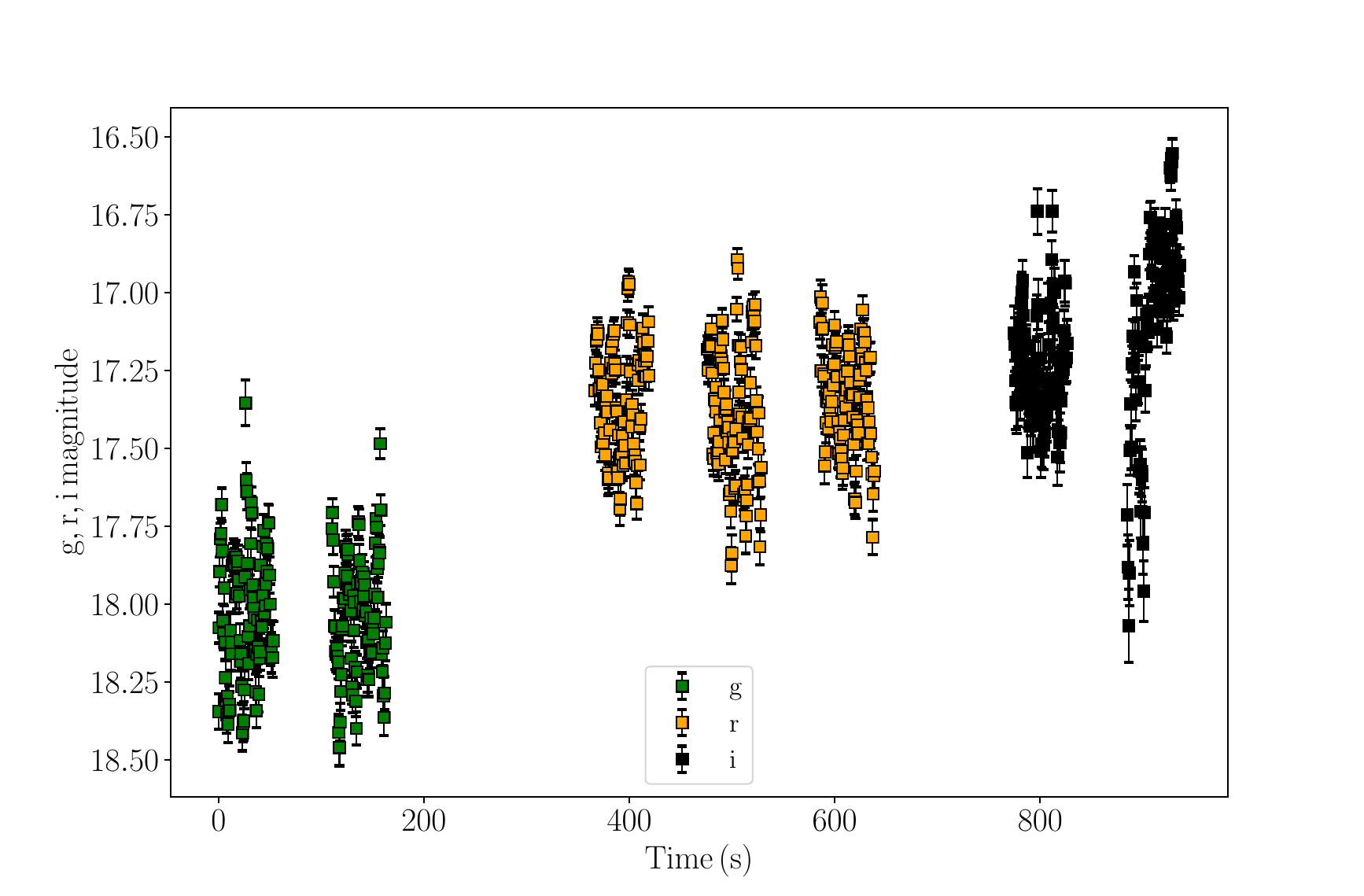}
\includegraphics[width=1\linewidth]{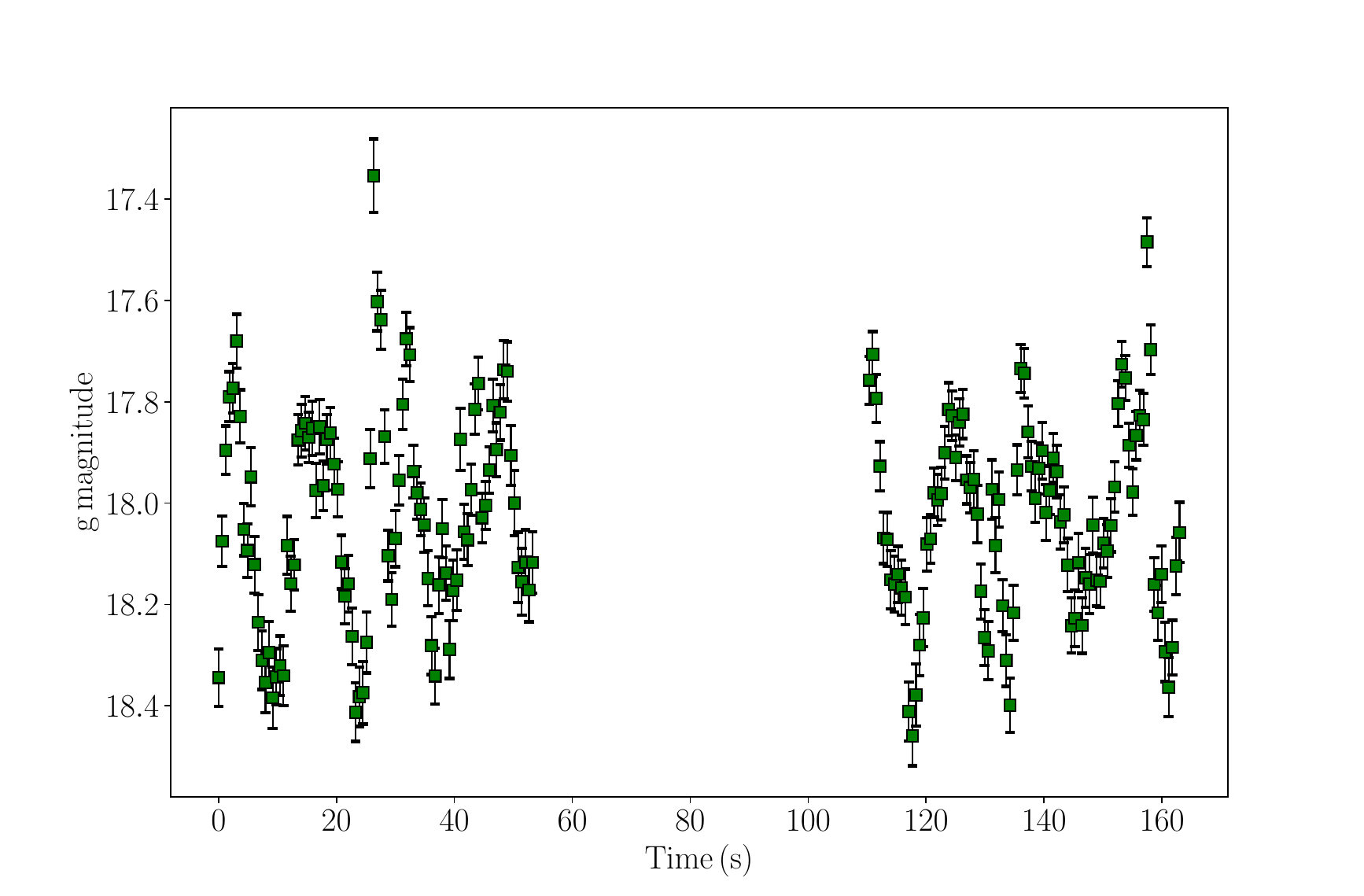}
\includegraphics[width=1\linewidth]{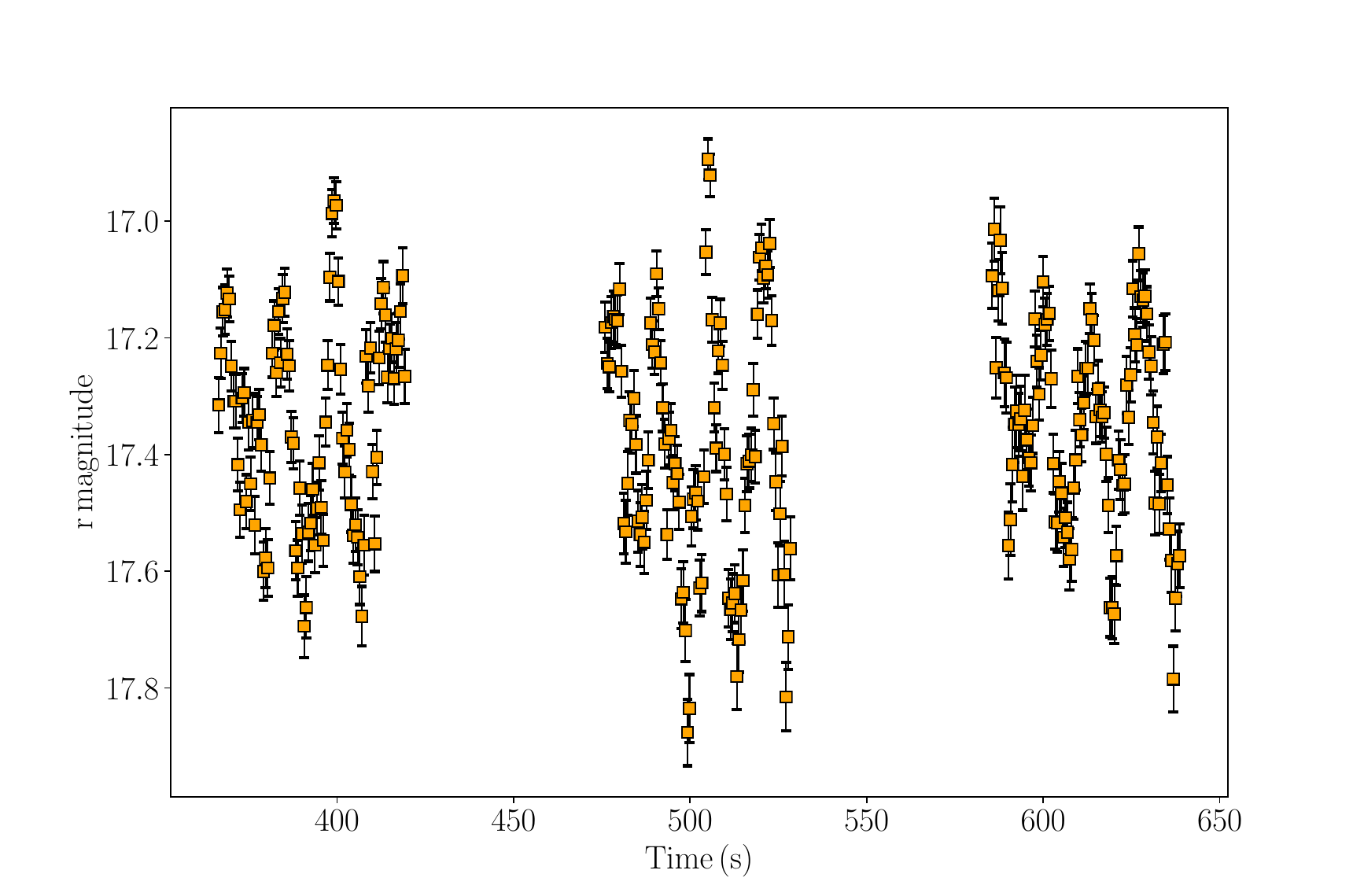}
\includegraphics[width=1\linewidth]{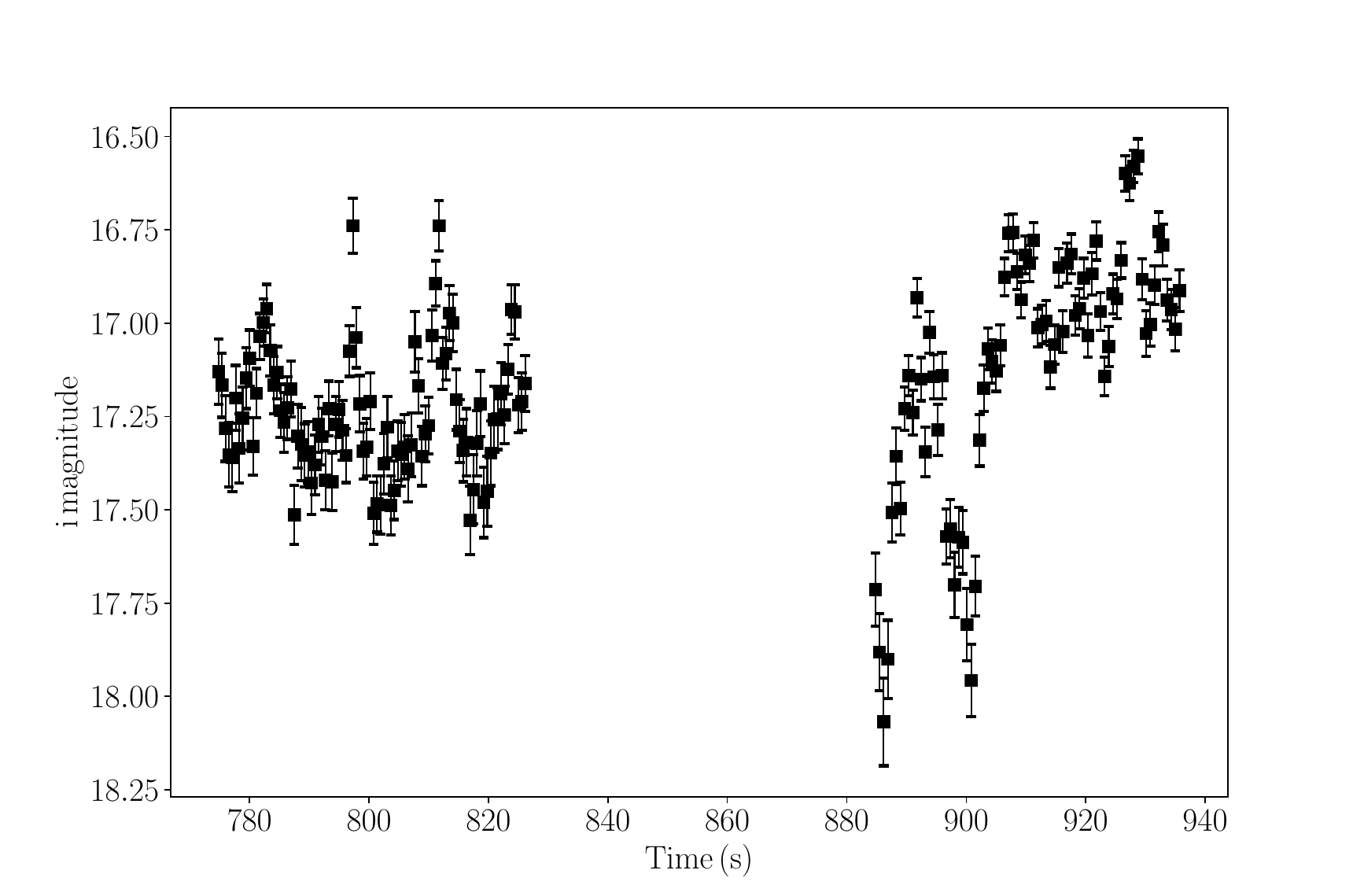}
\caption{Lightcurves in g, r and i bands from CFHT/MegaCam observations of \gea taken 2016 Apr 04. The total duration of the observations is $\sim$960~s.   First panel: combination all 3 bands. Second panel: g-band. Third panel: r-band. Fourth panel: i-band.}
\end{figure}

The combined g, r and i lightcurves for \cga (top panel of Fig.~S3) show clear evidence of the asteroid's colours (discussed below) while the individual g, r and i lightcurves (panels 2-4 of Fig.~S3) show an indication of a sinusoidal lightcurve shape that is more prominent in the r band images compared to the g and i band images.  The r band lightcurve has an amplitude of $\sim$0.6 magnitudes, but is diminished compared to the scatter in the g and i band lightcurves which could be due to tumbling motion or the lower SNR of the the g and i band data.

\begin{figure}
\centering 
\includegraphics[width=1\linewidth]{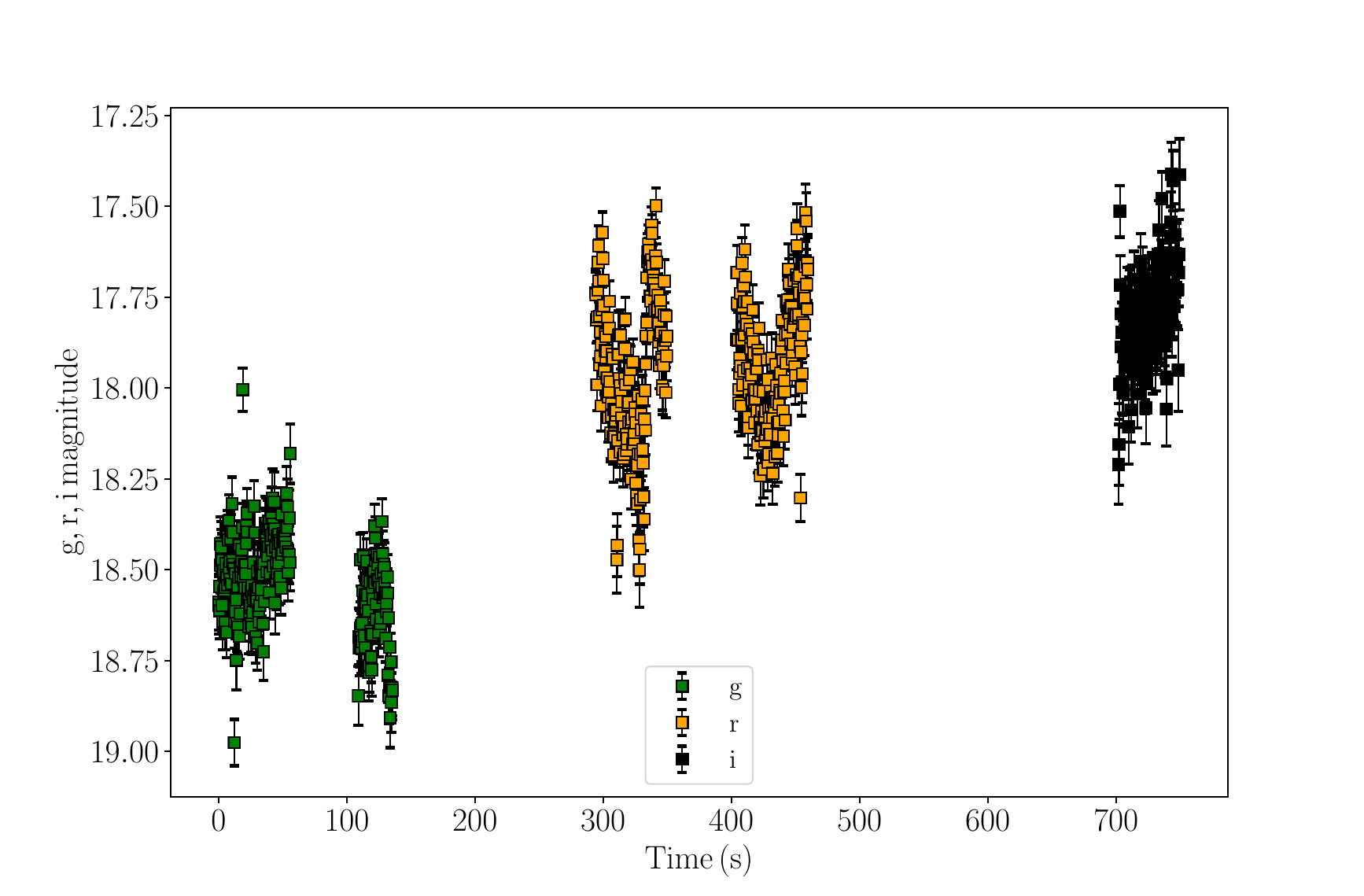}
\includegraphics[width=1\linewidth]{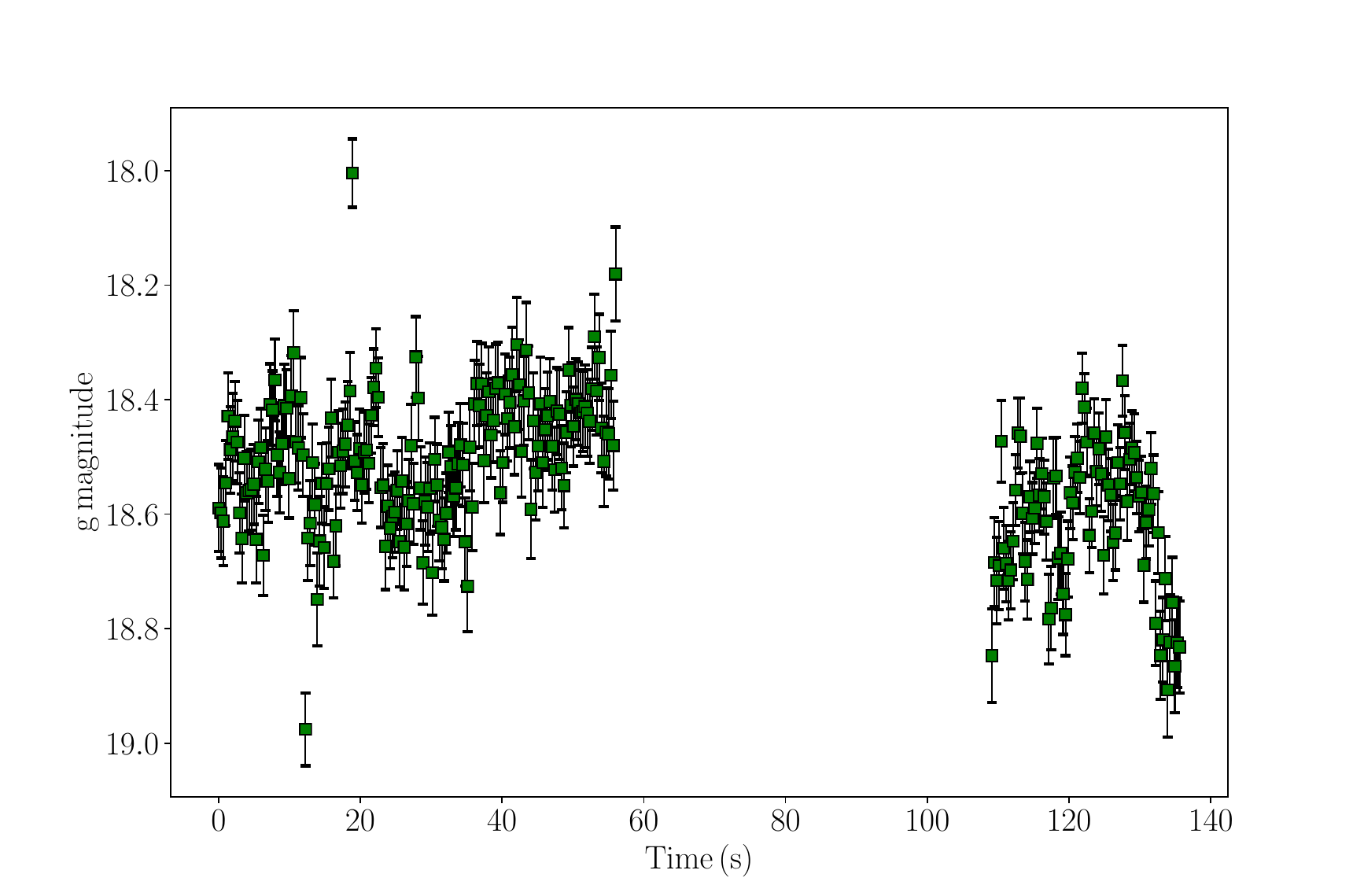}
\includegraphics[width=1\linewidth]{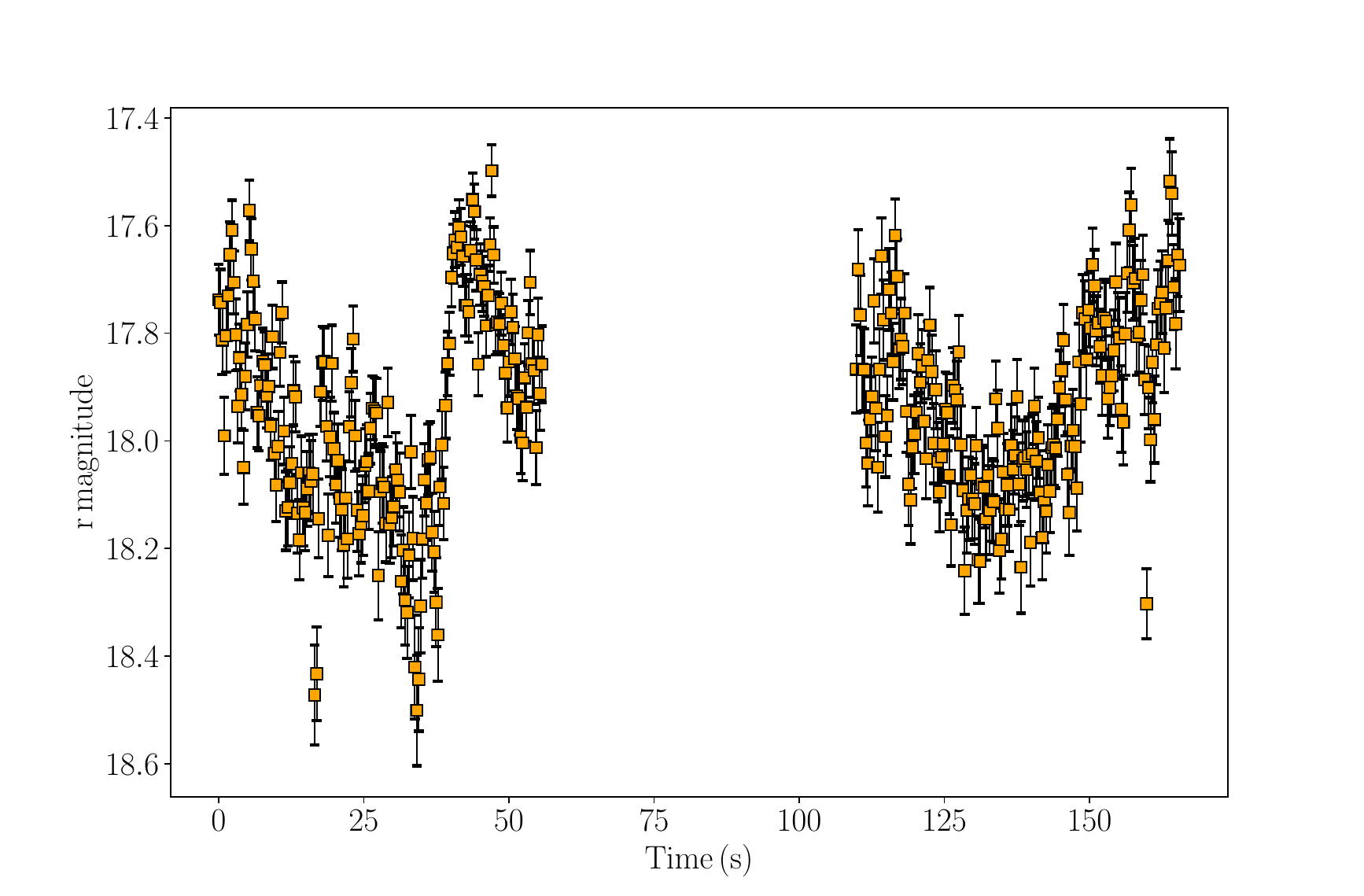}
\includegraphics[width=1\linewidth]{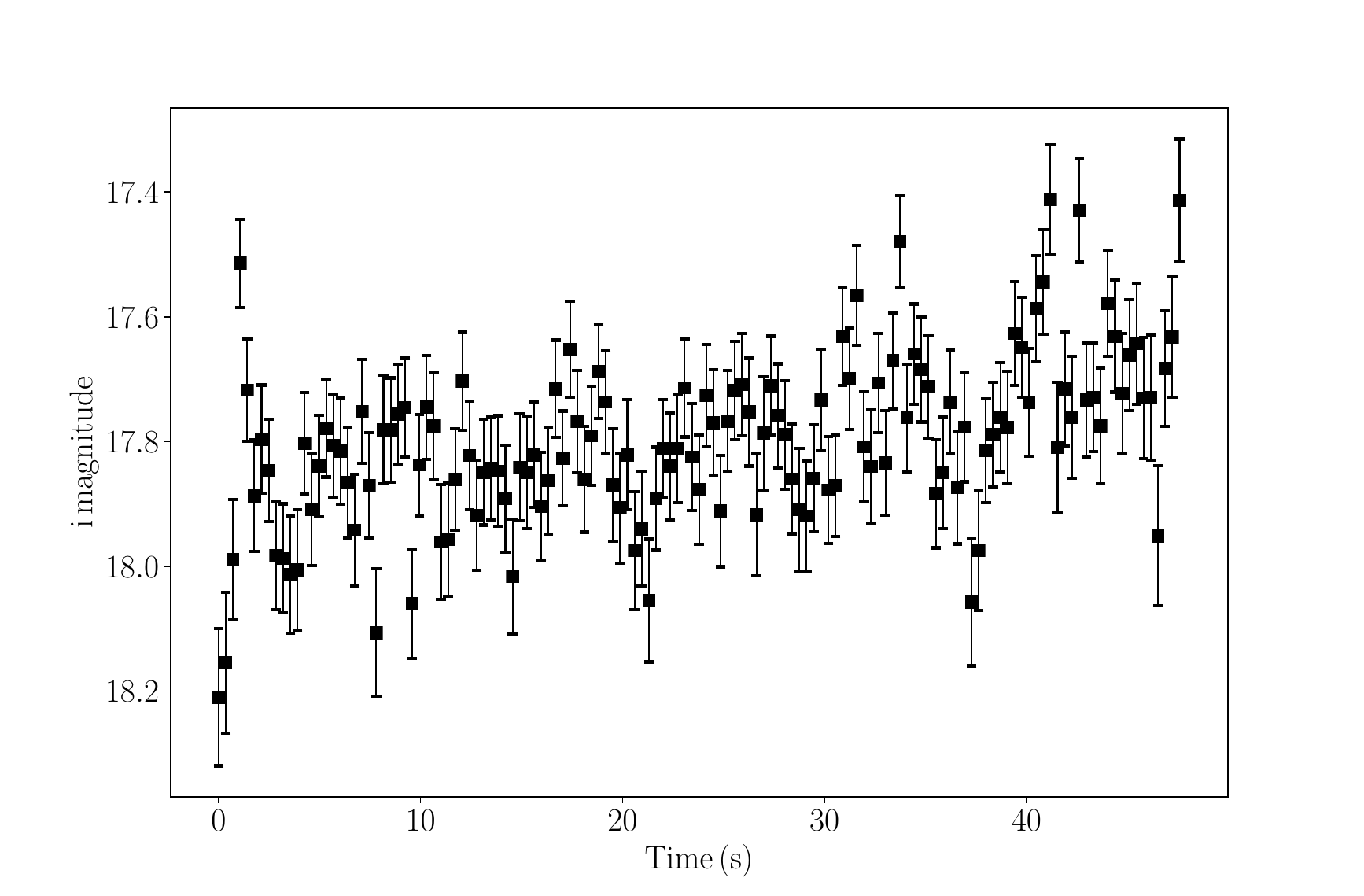}
\caption{
Lightcurves in g, r and i bands from CFHT/MegaCam observations of \cga taken 2016 Feb 05. The total duration of the observations is $\sim$750~s.   First panel: combination all 3 bands. Second panel: g-band. Third panel: r-band. Fourth panel: i-band.}
\end{figure}

The combined g, r and i lightcurves for \eva (top panel of Fig.~S4) again shows evidence of the asteroid's colours (discussed below) while the individual g and r lightcurves (panels 2-3 of Fig.~S4) show a sinusoidal shape with an ampltude of $\sim$0.9 mags. The i band lightcurve (panels4 of Fig.~S4) shows a hint of a sinusoidal shape but an amplitude comparable to the scatter in the lightcurve.

\begin{figure}
\centering 
\includegraphics[width=1\linewidth]{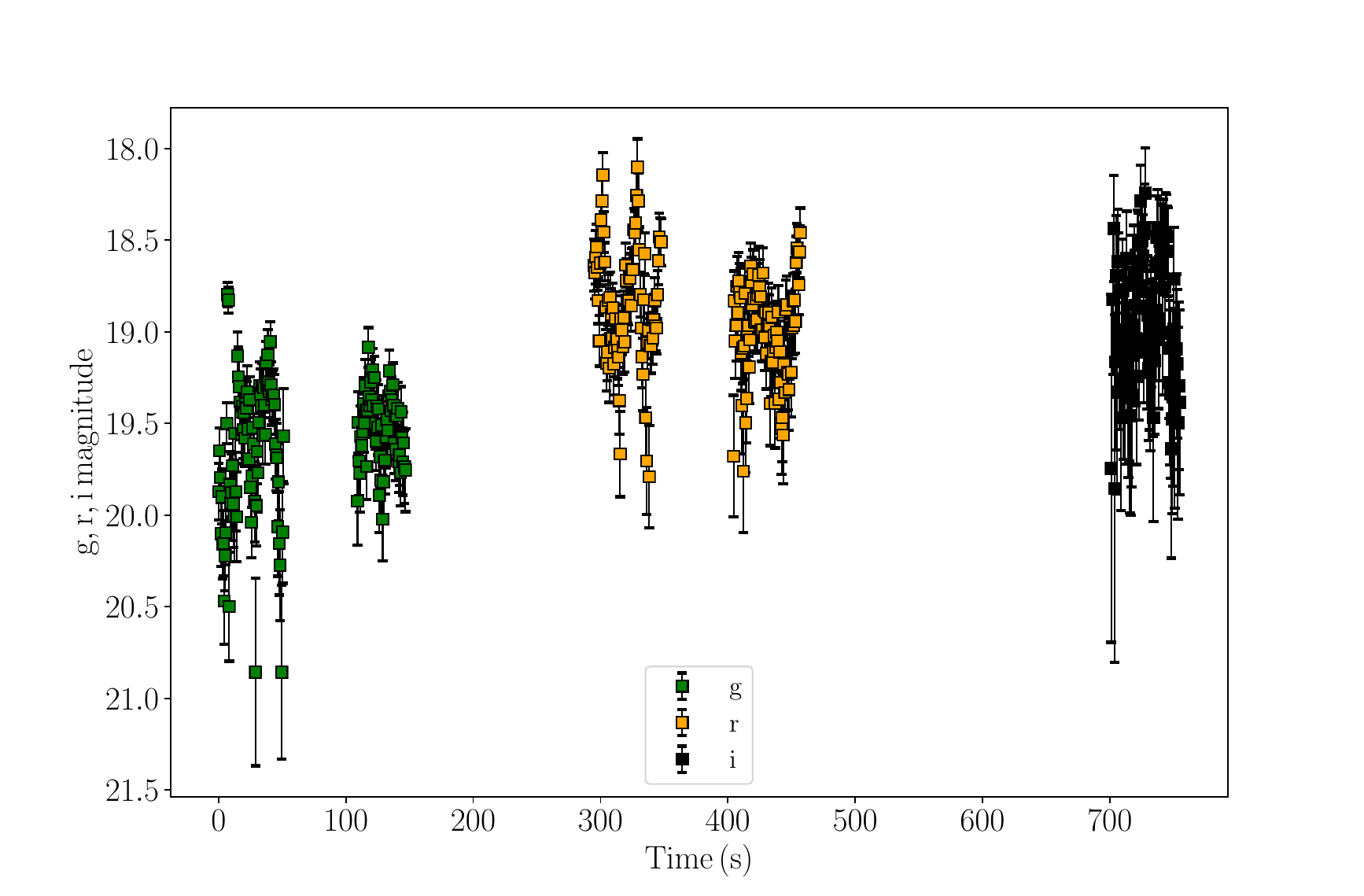}
\includegraphics[width=1\linewidth]{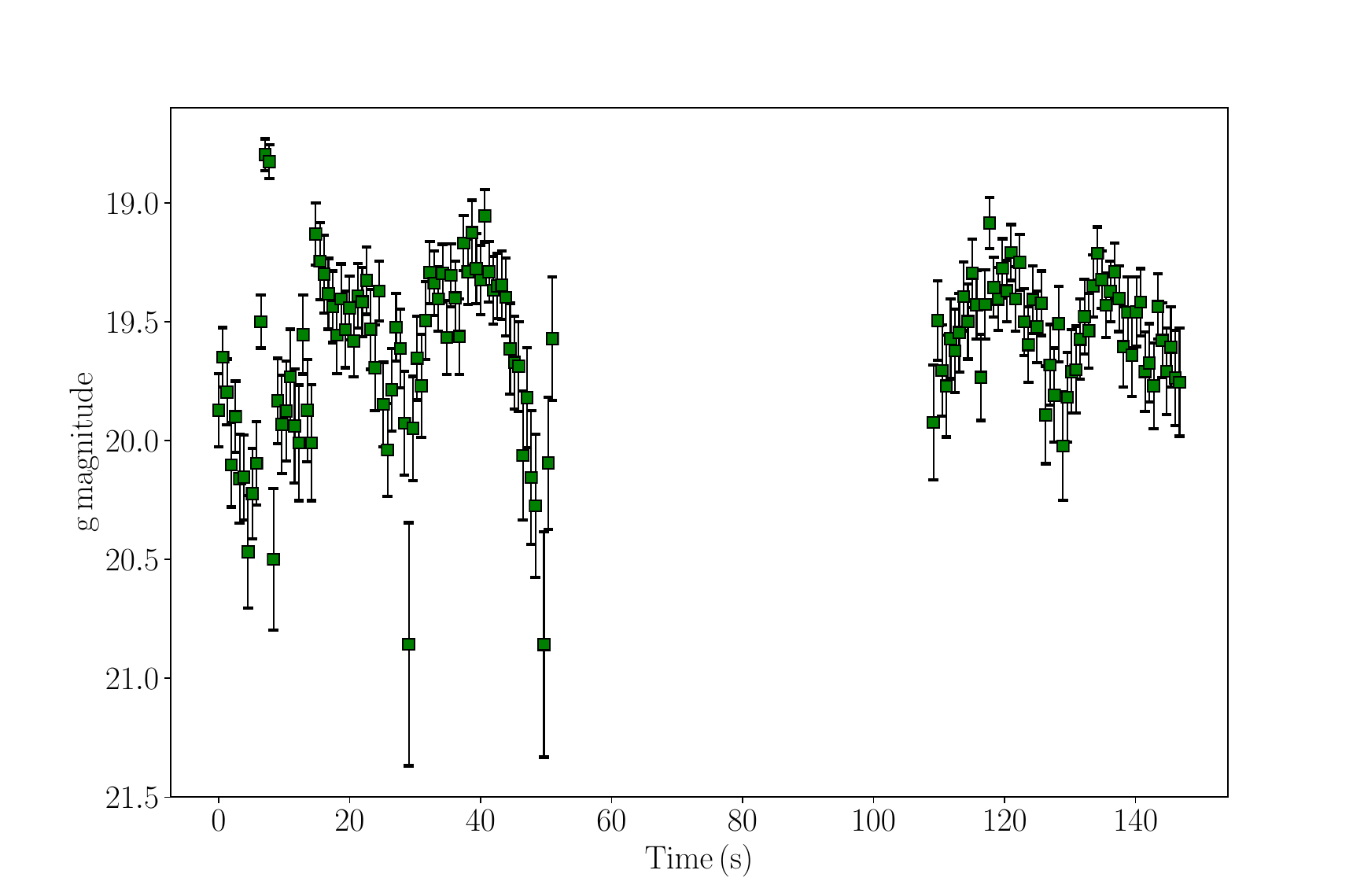}
\includegraphics[width=1\linewidth]{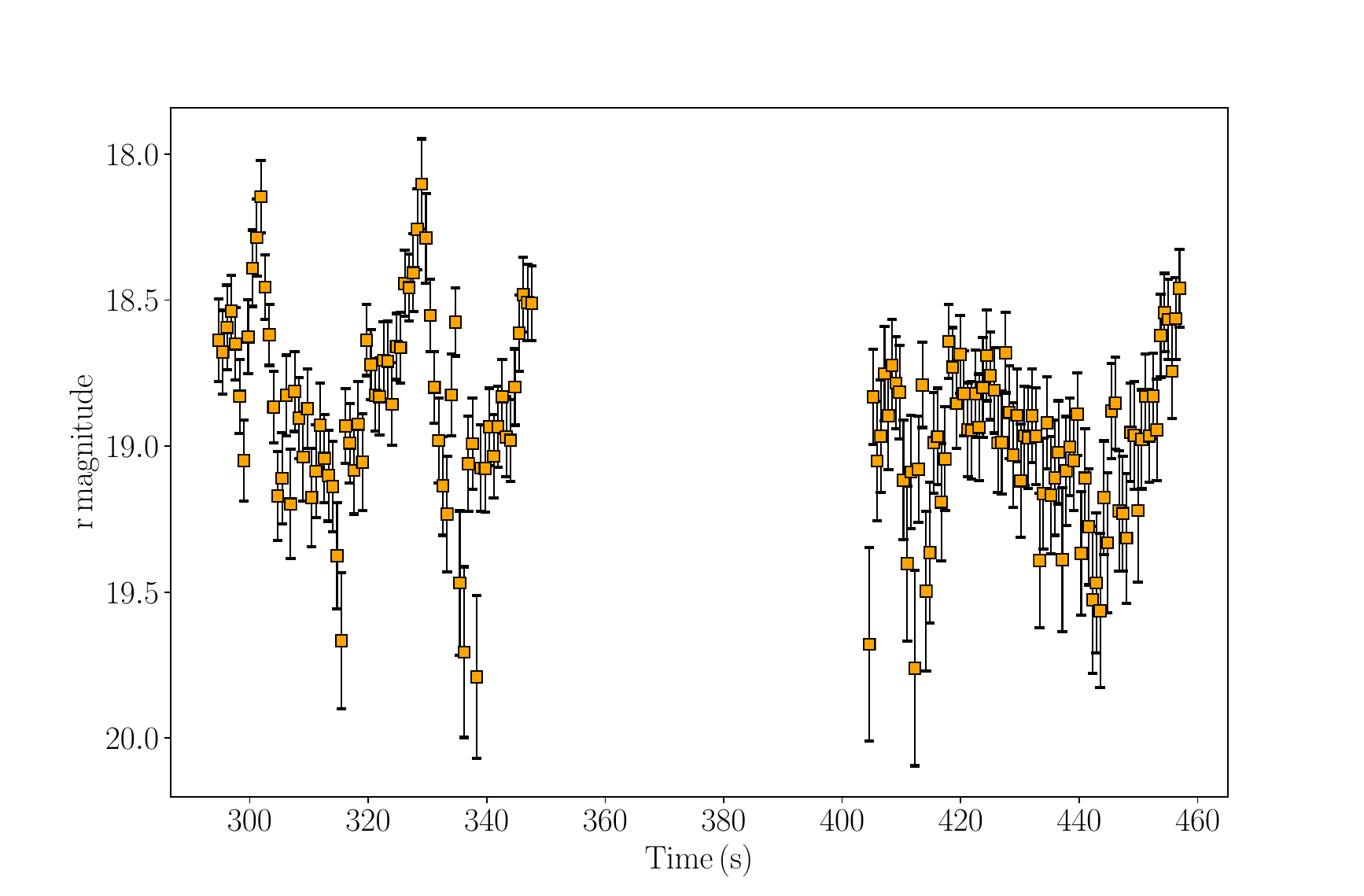}
\includegraphics[width=1\linewidth]{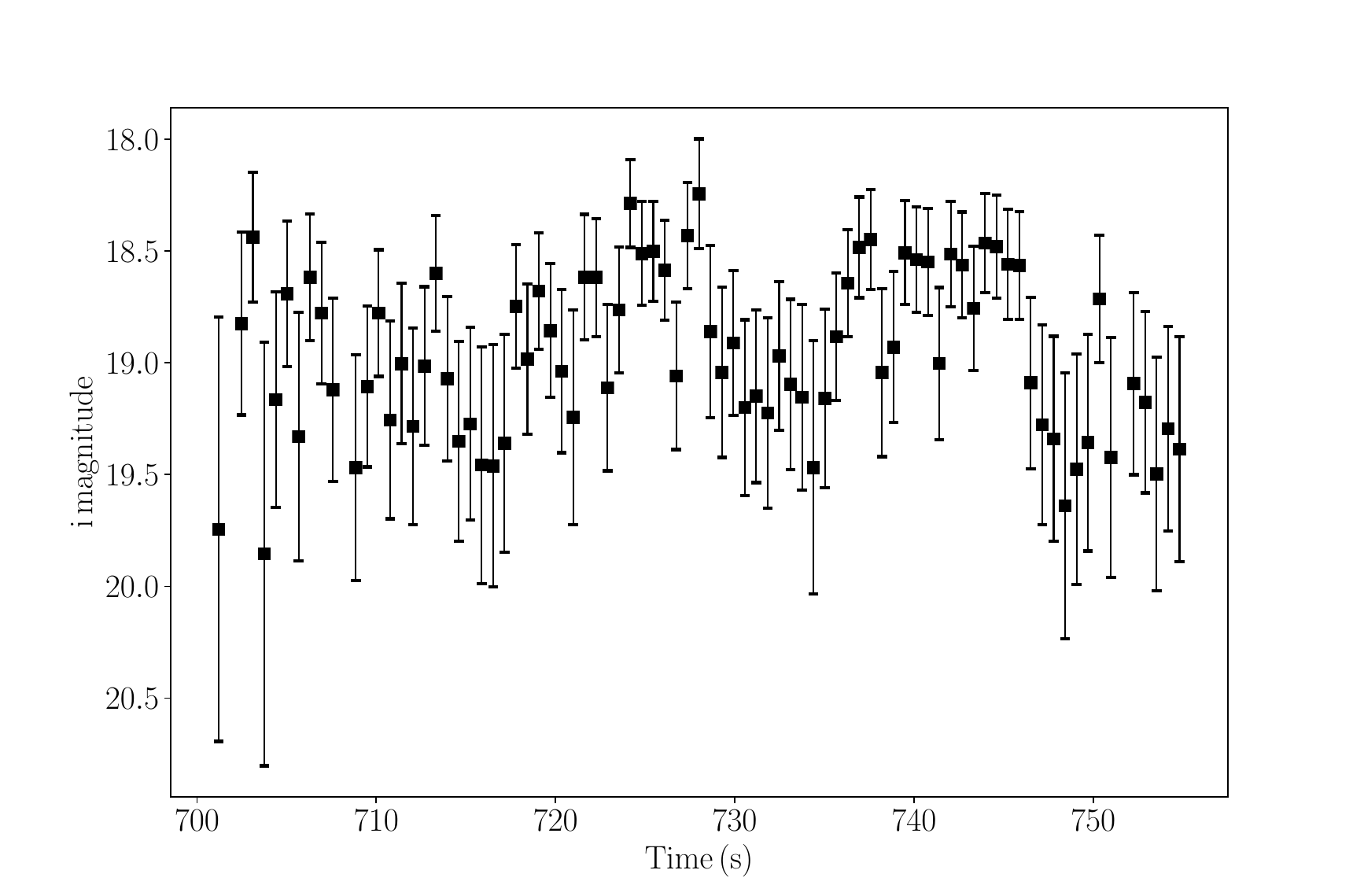}
\caption{
Lightcurves in g, r and i bands from CFHT/MegaCam observations of \eva taken 2016 Mar 12. The total duration of the observations is $\sim$755~s.   First panel: combination all 3 bands. Second panel: g-band. Third panel: r-band. Fourth panel: i-band.}
\end{figure}

\subsection{Periods and colours}

We applied the Lomb-Scargle periodogram (LS) \citep[][]{Lomb1976,Scargle1982} to the combined, detrended g, r and i lightcurve data to determine the rotation period of our three asteroids (top panels of Figs.~S5-7). The maximum period in the LS periodogram search was the total time span of the observations (column 13 in Table~S1). For \cgans, the maximum period in the LS periodogram search was the total duration of the r band images ($\sim$140~s) which is much more than the found lightcurve period ($\sim$28~s, top panel of Fig.~S6). We applied bootstrap estimation of the uncertainties \citep[][]{Press1986} by randomly removing $\sqrt{N}$ data points from the time-series lightcurves and repeating the periodogram estimation of the lightcurve period 1,000 times resulting in a 1-$\sigma$ estimate of the lightcurve rotation period uncertainties of 0.01-0.1 s. The g, r and i lightcurves were then folded assuming a double-peak rotational lightcurve (second panels of Figs.~S5-7) and re-binned with bin widths of 0.05 for phases in the range 0-1 (third panel of Figs.~S5-7).  Finally, the g-r and r-i colour differences were calculated as a function of phase (bottom panels of Figs.~S5-7).

\geans's LS power spectrum peak indicates a lightcurve period of $\sim$15~s (top panel of Fig.~S5) implying a double-peak rotation period of $\sim$31~s that was used for folding the lightcurve data (second and third panel of Fig.~S5). The weighted mean g-r colour for \eva is $\sim$0.66$\pm$0.01 and the weighted mean r-i colours is $\sim0.20\pm0.01$ (fourth panel of Fig.~S5).

\begin{figure}\centering
\hspace{0 mm}
\centering
\includegraphics[width=1\linewidth]{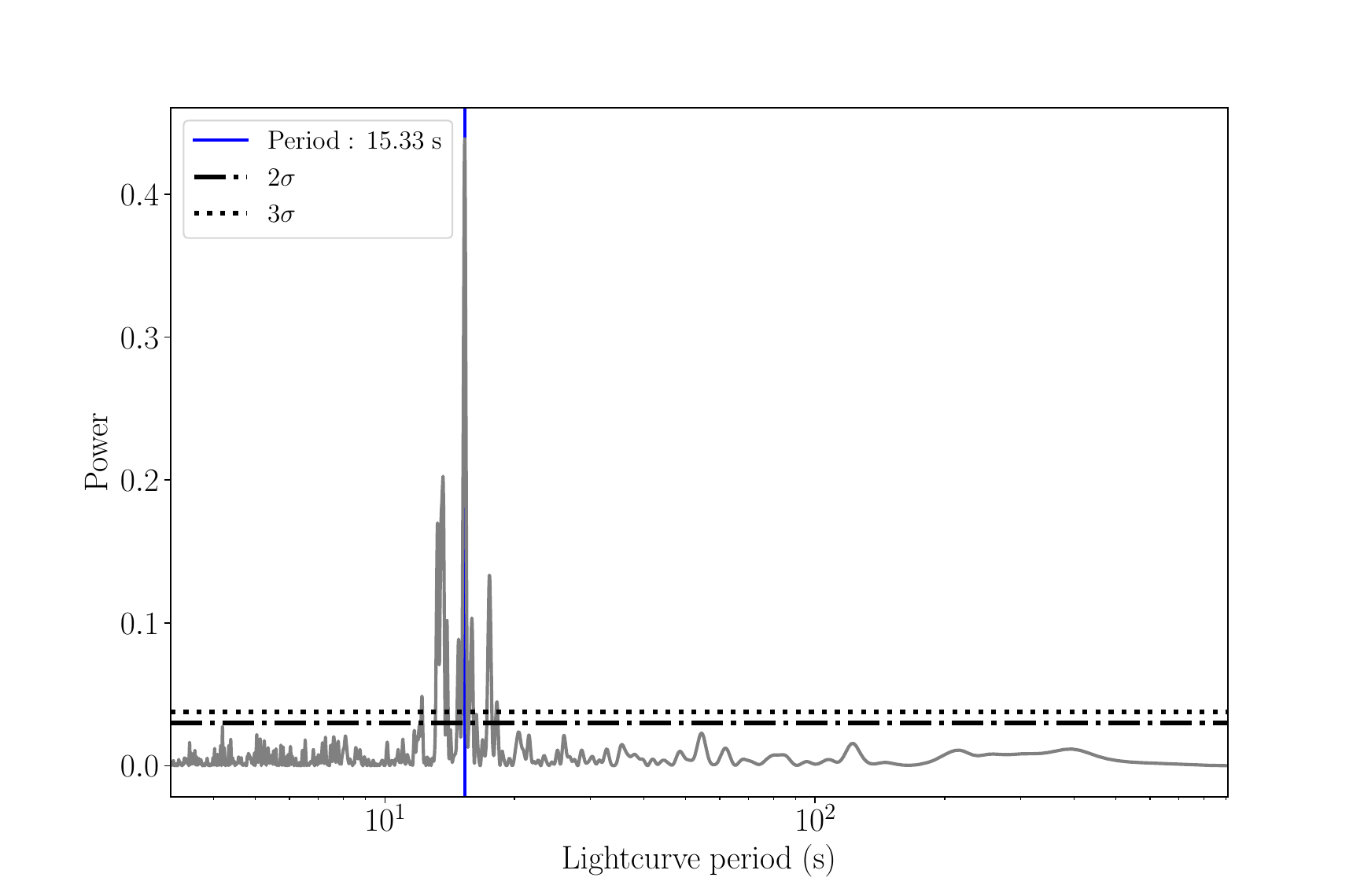}
\includegraphics[width=1\linewidth]{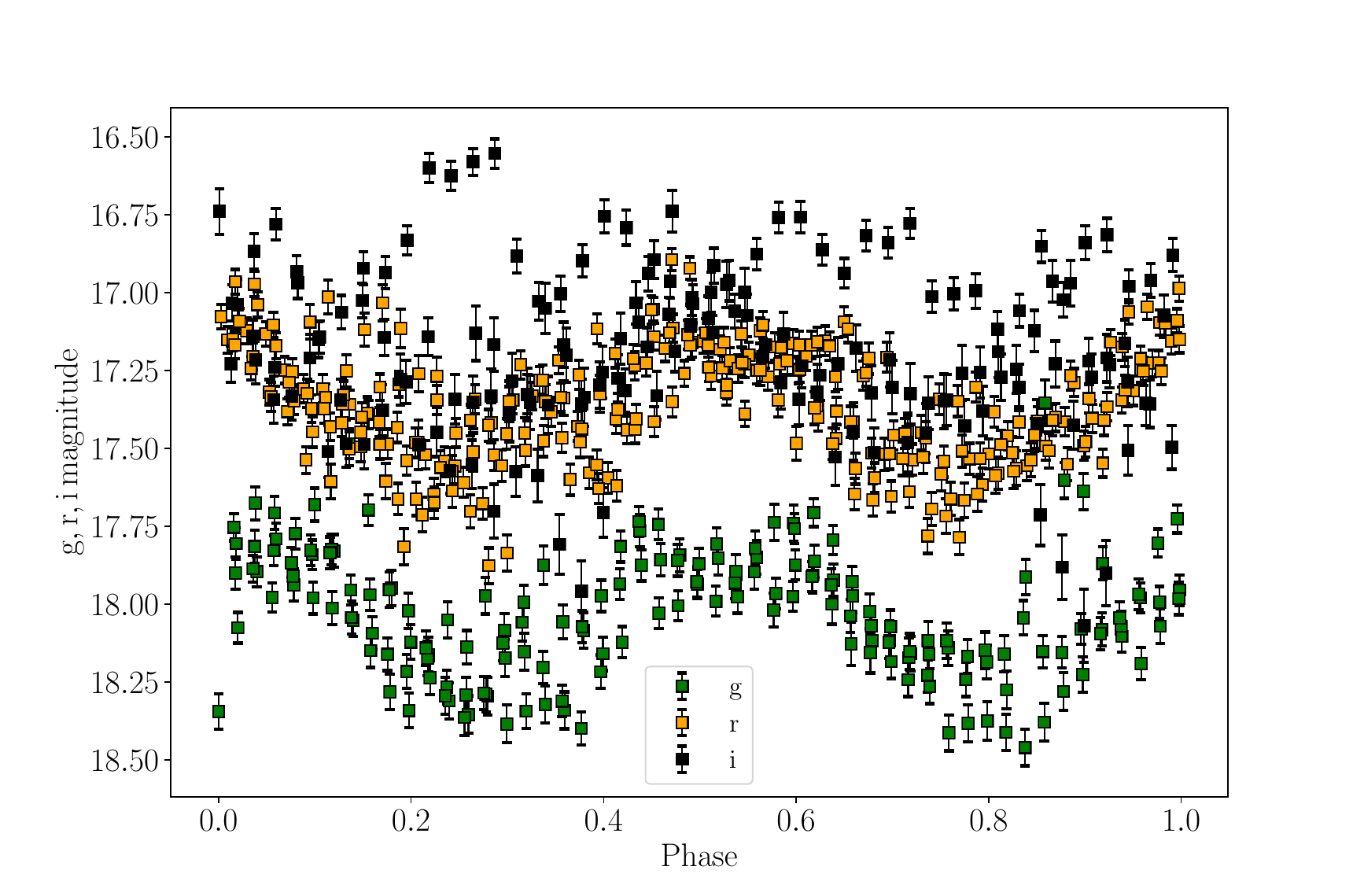}
\includegraphics[width=1\linewidth]{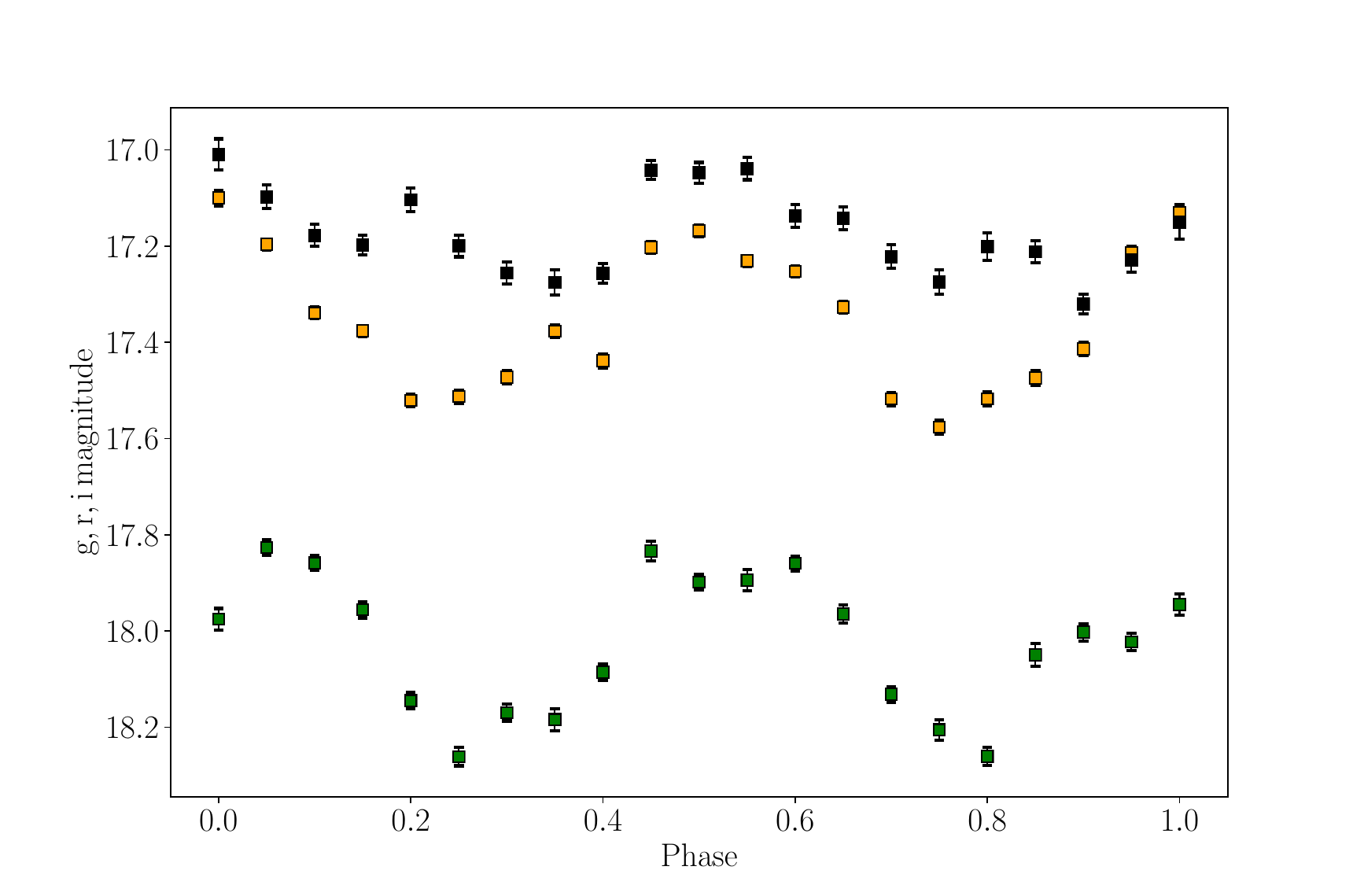}
\includegraphics[width=1\linewidth]{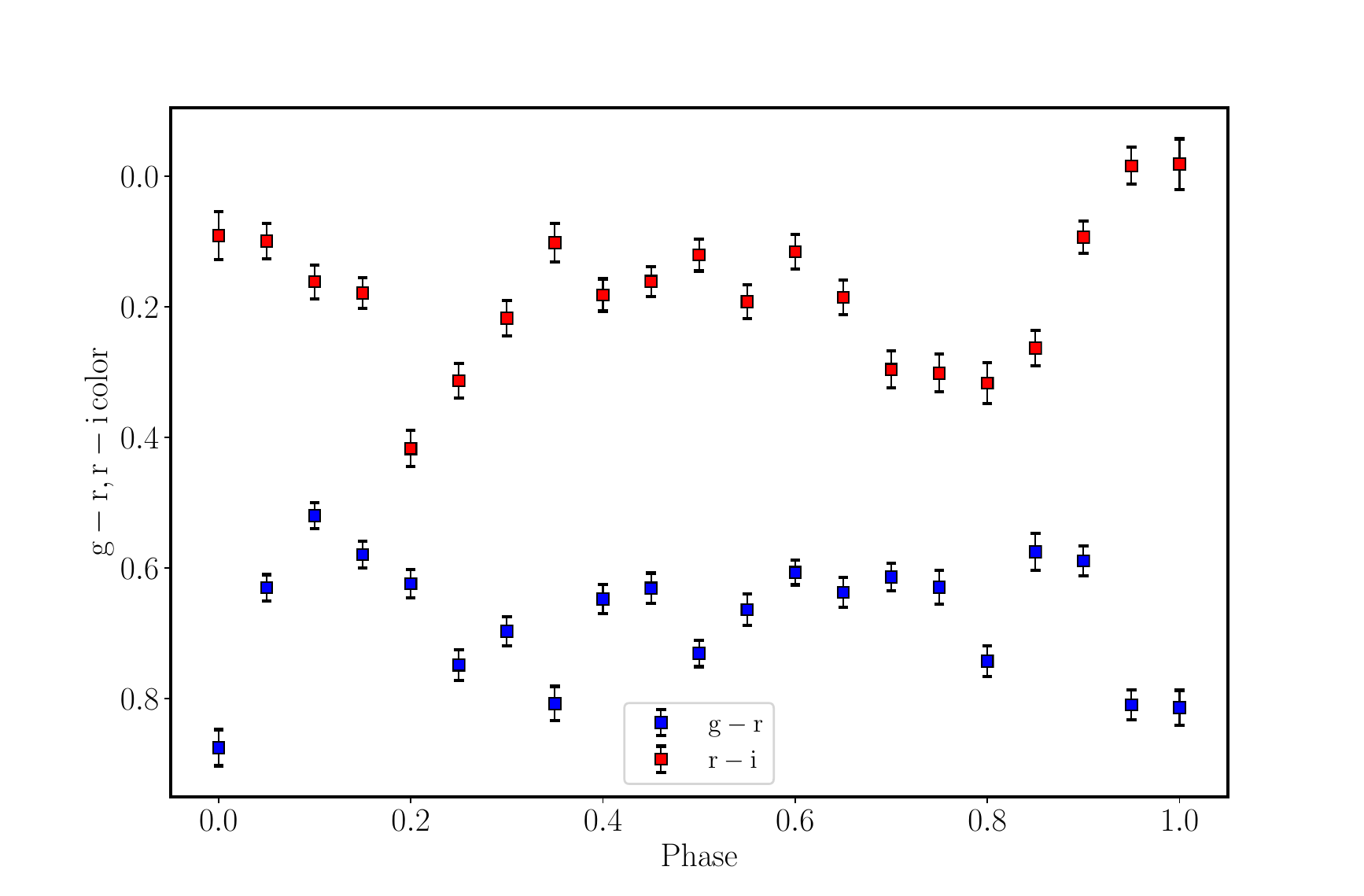}
\caption{Top panel: Lomb-Scargle periodogram of the r-band data for \geans. The 2-$\sigma$ and 3-$\sigma$ false alarm probabilities are shown as dotted and dashed-dotted lines respectively. A local maxima of $\sim$15.3~s is indicated with a vertical blue line. Second panel: the g, r and i-band lightcurve data of \gea folded with a rotation period of $\sim$31~s. Third panel: The g, r and i lightcurve data of \gea folded to a rotation period of $\sim$31~s and rebinned with a phase bin size of 0.05. Fourth panel: g-r and r-i rebinned and folded colour curves of \gea.}
\end{figure}

\cgans's LS power spectrum peak indicates a lightcurve period of $\sim$28~s (top panel of Fig.~S6) implying a double-peak rotation period of $\sim$56~s that was used for folding the lightcurve data (second and third panel of Fig.~S6). The weighted mean g-r colour for \eva is $\sim$0.58$\pm$0.01 and the weighted mean r-i colours is $\sim0.17\pm0.01$ (fourth panel of Fig.~S6).

\begin{figure}
\centering 
\includegraphics[width=1\linewidth]{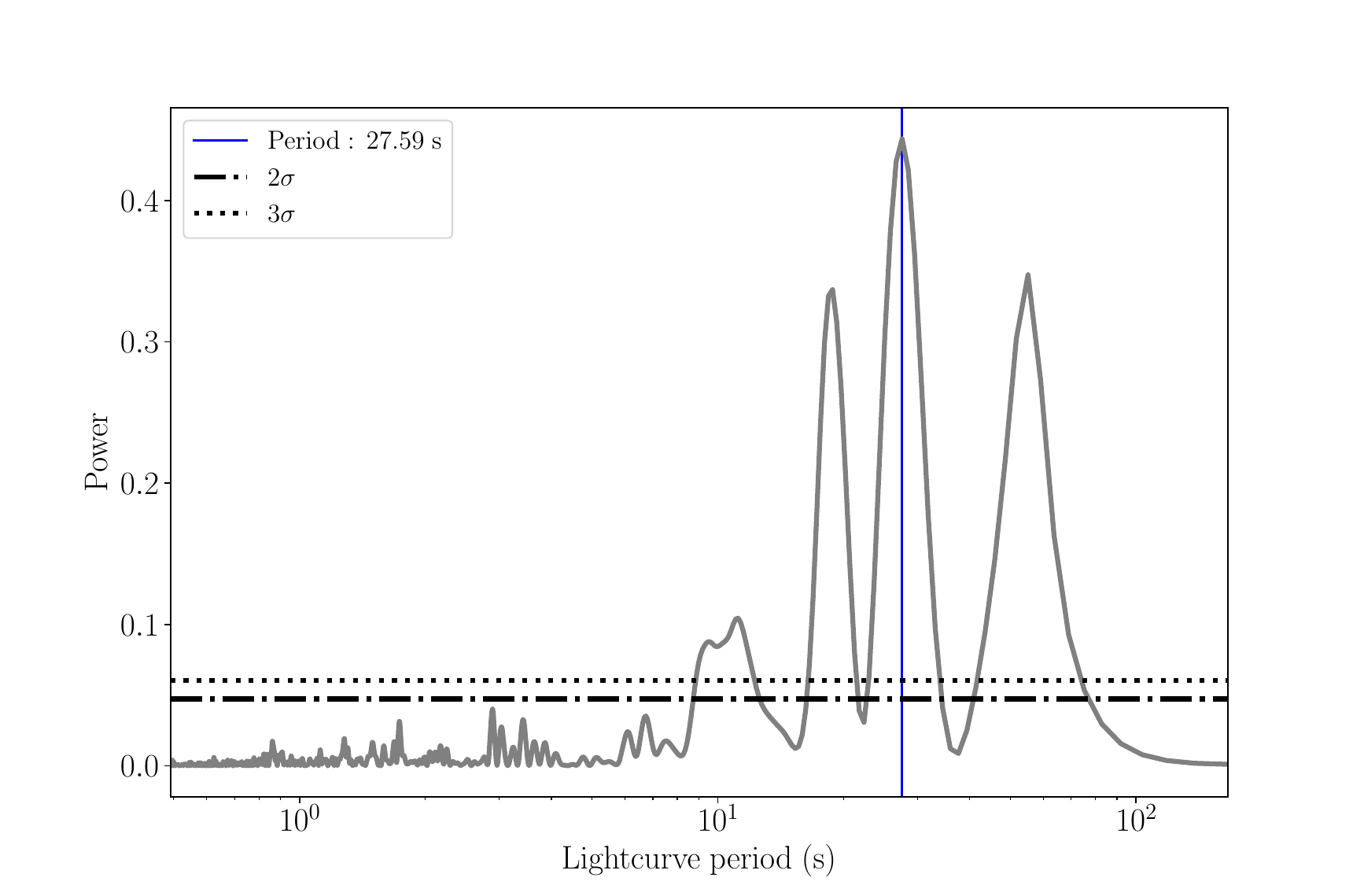}
\includegraphics[width=1\linewidth]{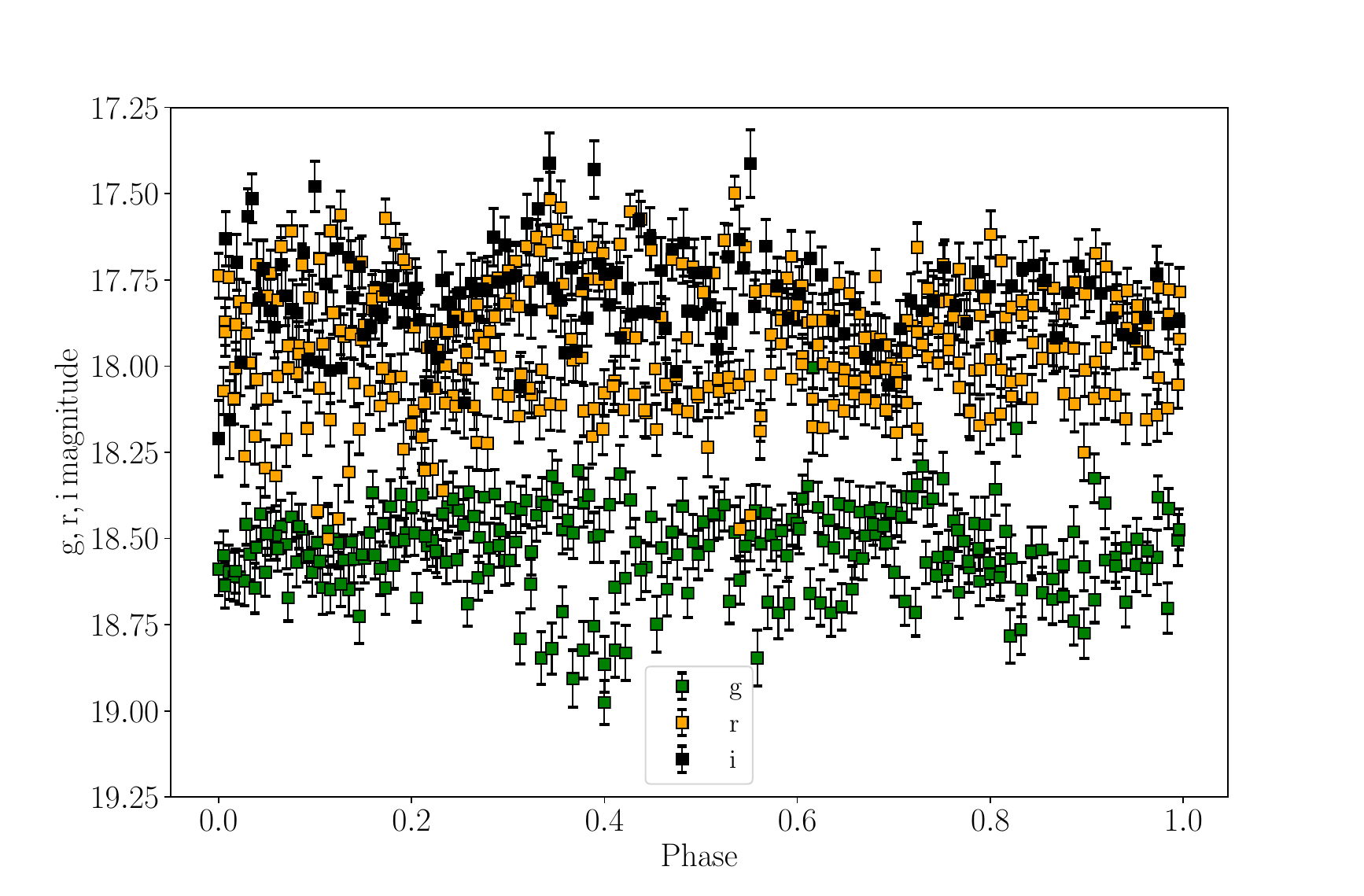}
\includegraphics[width=1\linewidth]{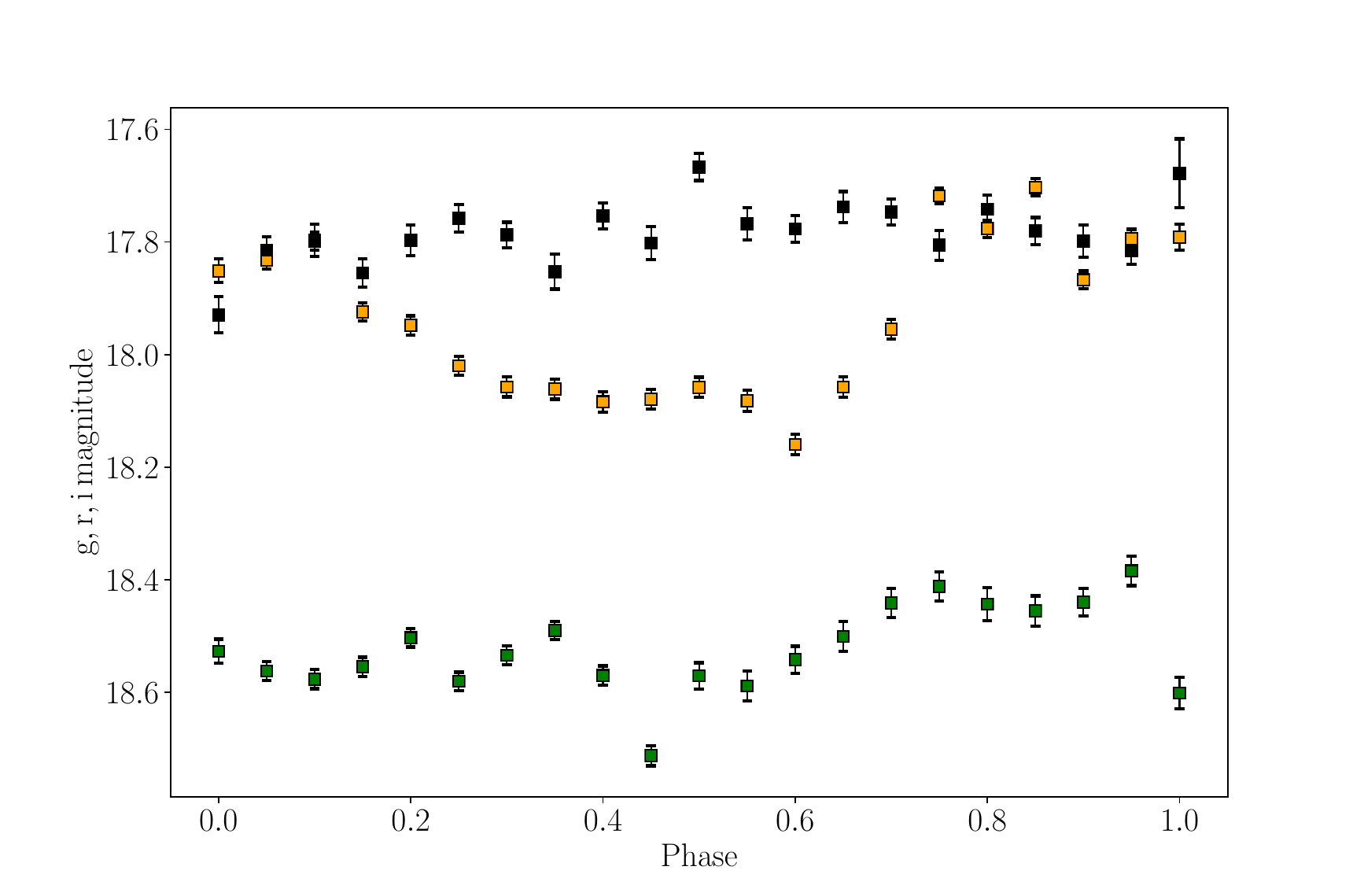}
\includegraphics[width=1\linewidth]{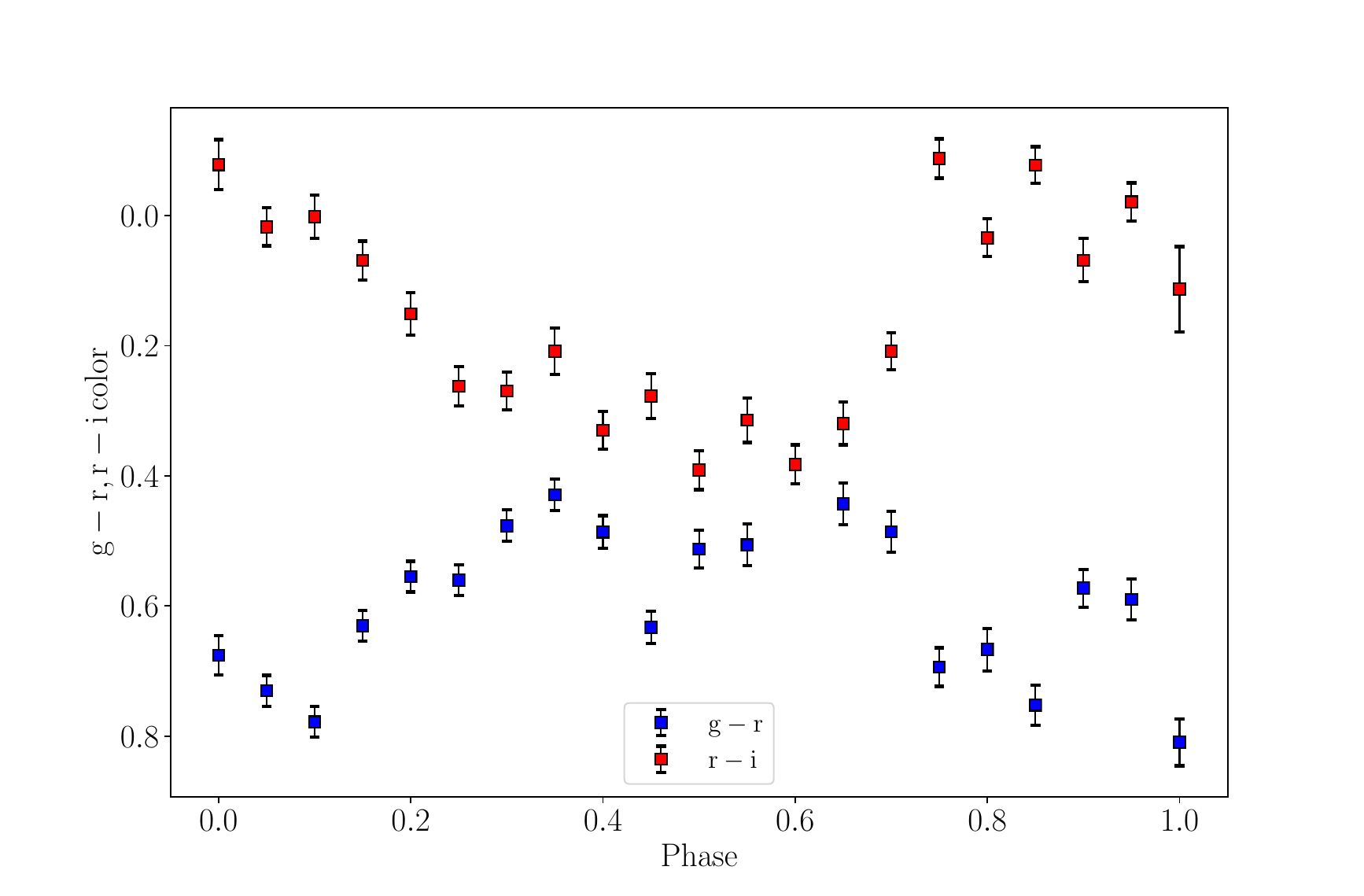}
\caption{Top panel: Lomb-Scargle periodogram of the r-band data for \cgans. The 2-$\sigma$ and 3-$\sigma$ false alarm probabilities are shown as dotted and dashed-dotted lines respectively. A local maxima of $\sim$28~s is indicated with a vertical blue line. Second panel: the g, r and i-band lightcurve data of \cga folded with a rotation period of $\sim$56~s. Third panel: The g, r and i lightcurve data of \cga folded to a rotation period of $\sim$56~s and rebinned with a phase bin size of 0.05. Fourth panel: g-r and r-i rebinned and folded colour curves of \cga.}
\end{figure}

\evans's LS power spectrum peak indicates a lightcurve period of $\sim$26~s (top panel of Fig.~S7) implying a double-peak rotation period of $\sim$52~s that was used for folding the lightcurve data (second and third panel of Fig.~S7). The weighted mean g-r colour for \eva is $\sim$0.70$\pm$0.02 and the weighted mean r-i colours is $\sim0.05\pm0.04$ (fourth panel of Fig.~S7).

\begin{figure}
\centering 
\includegraphics[width=1\linewidth]{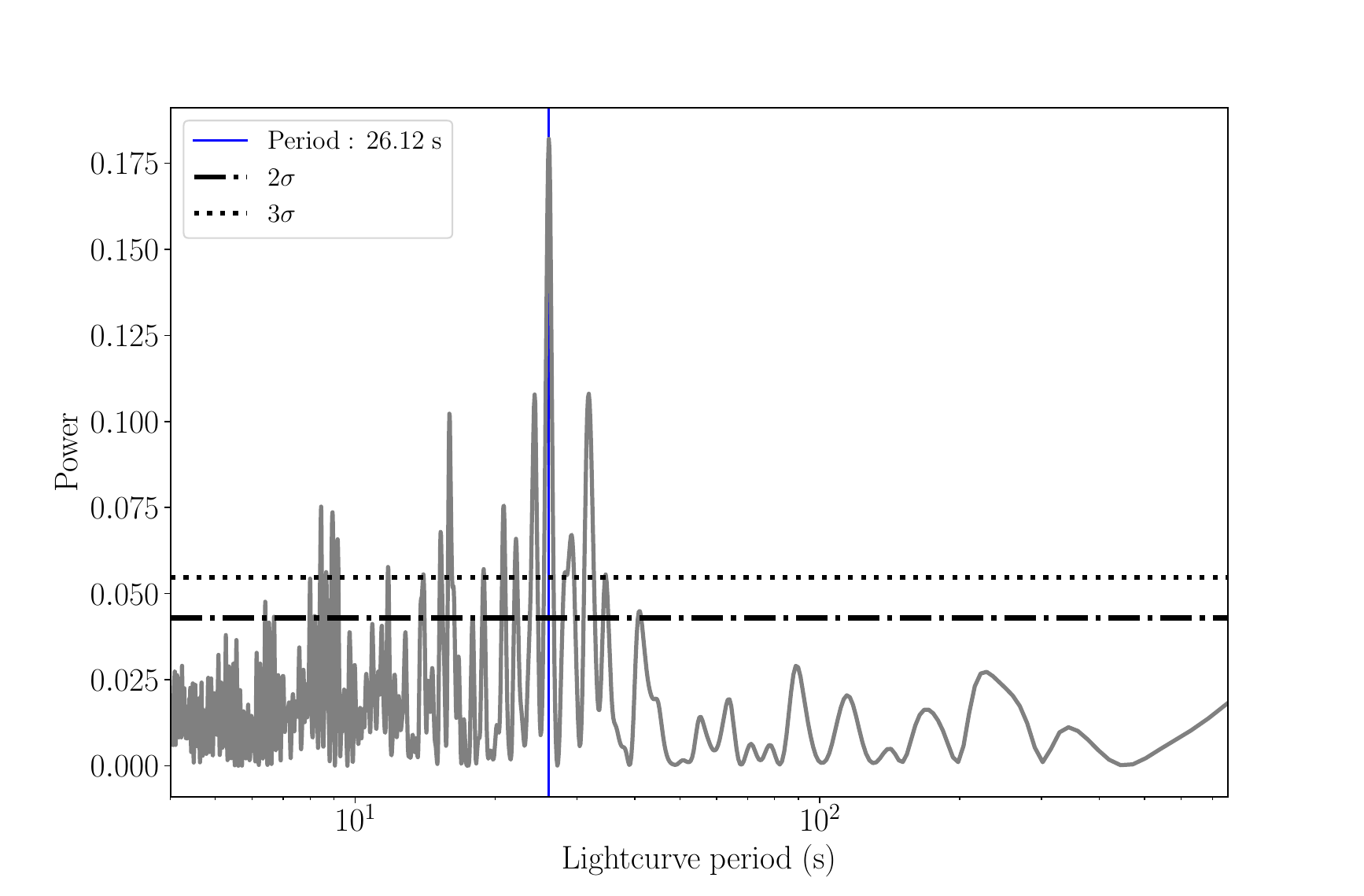}
\includegraphics[width=1\linewidth]{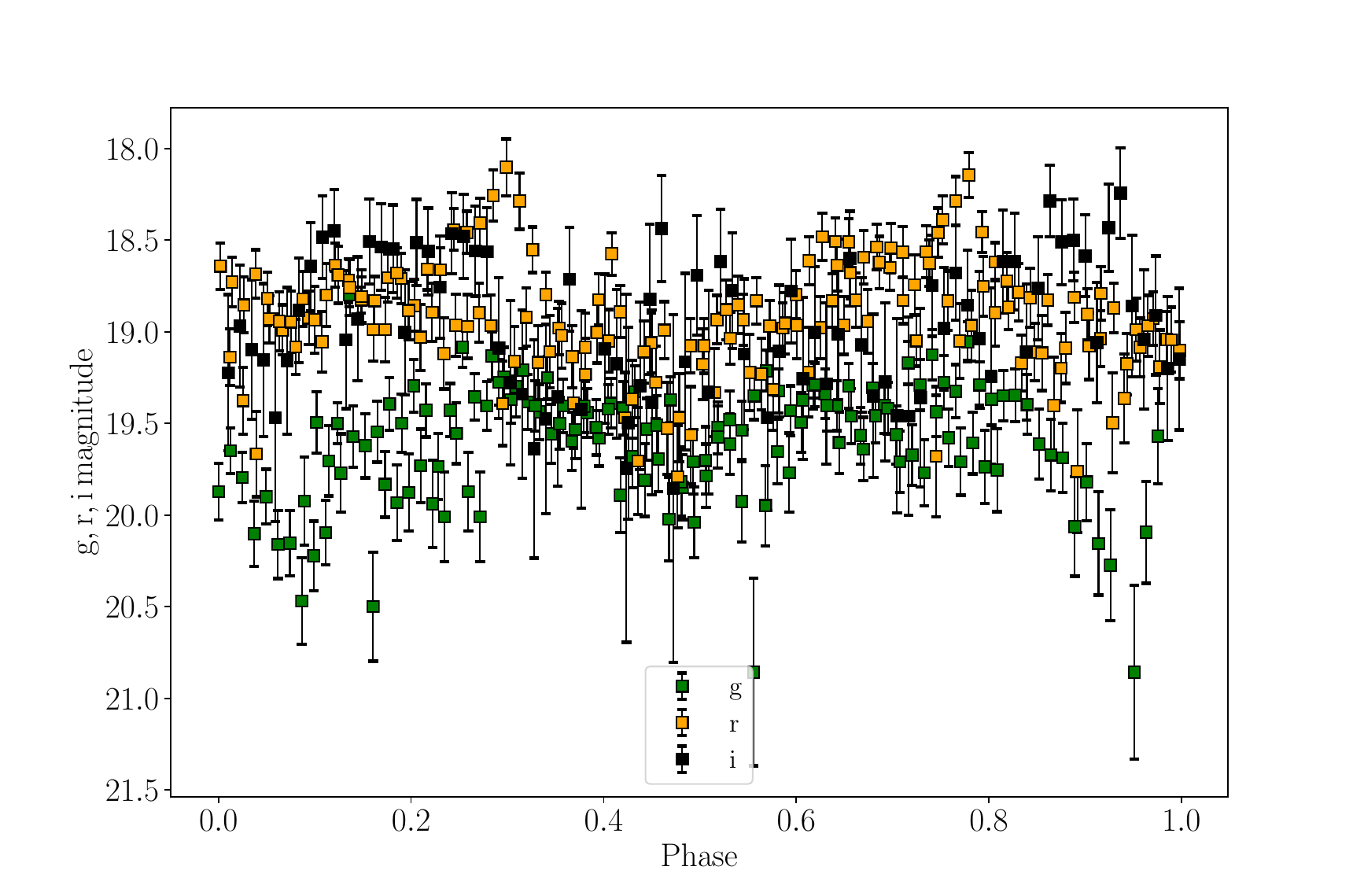}
\includegraphics[width=1\linewidth]{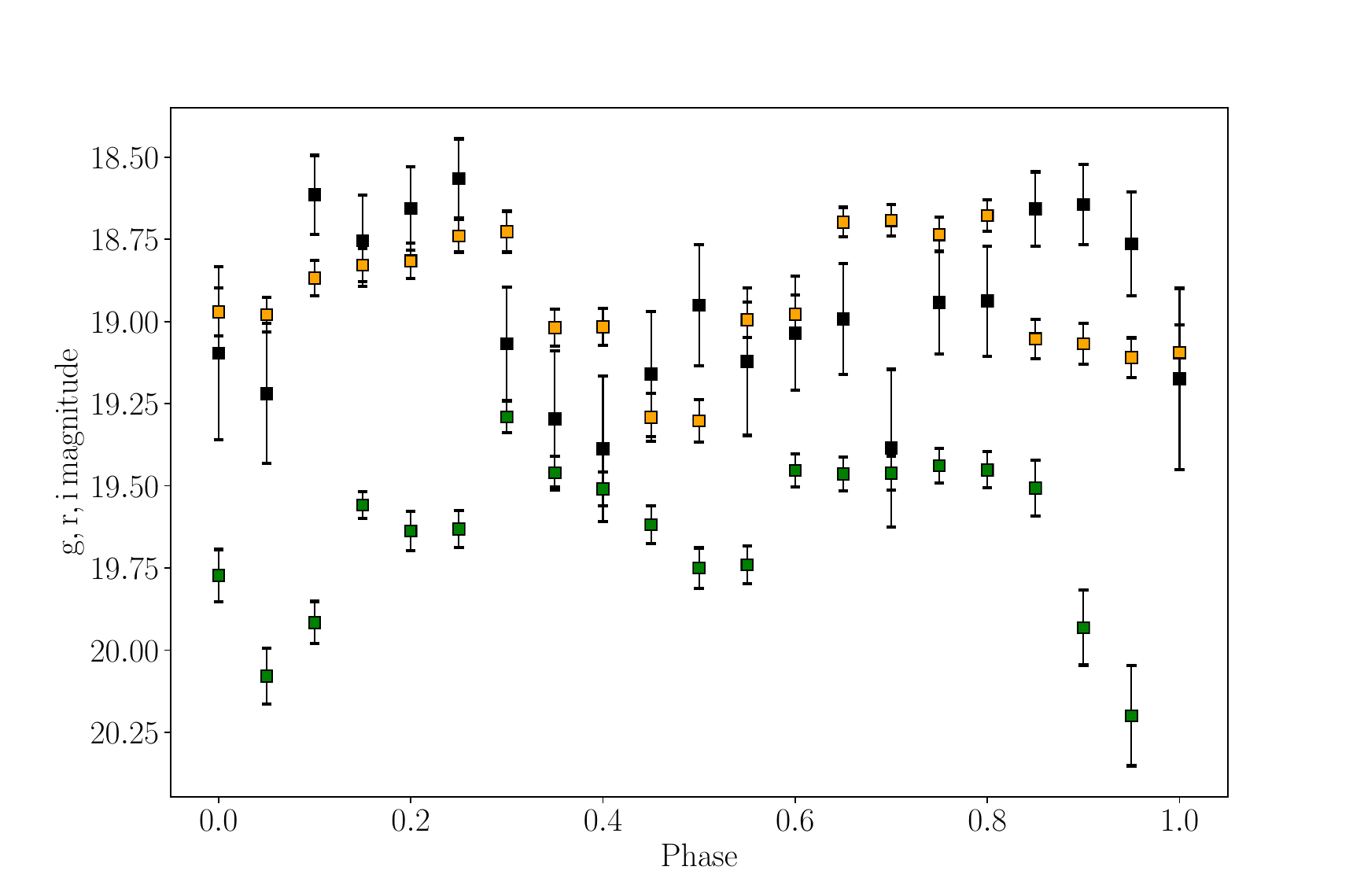}
\includegraphics[width=1\linewidth]{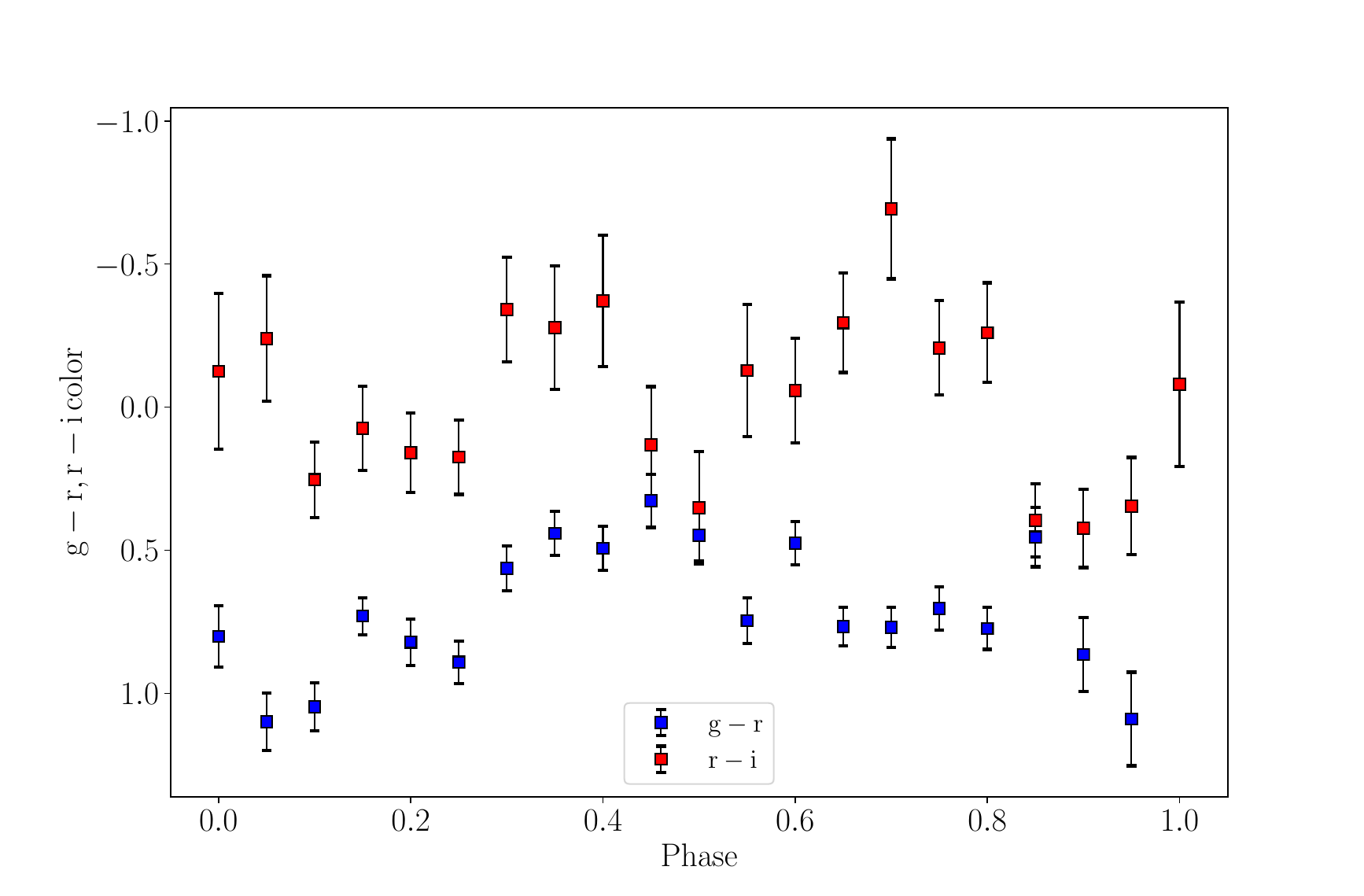}
\caption{Top panel: Lomb-Scargle periodogram of the r-band data for \evans. The 2-$\sigma$ and 3-$\sigma$ false alarm probabilities are shown as dotted and dashed-dotted lines respectively. A local maxima of $\sim$26~s is indicated with a vertical blue line. Second panel: the g, r and i-band lightcurve data of \eva folded with a rotation period of $\sim$52 s. Third panel: The g, r and i lightcurve data of \eva folded to a rotation period of $\sim$52~s and rebinned with a phase bin size of 0.05. Fourth panel: g-r and r-i rebinned and folded colour curves of \eva.}
\end{figure}

\begin{landscape}
\begin{table}\vphantom{13pt}\vphantom{43pt}\vphantom{13pt}\vphantom{13pt}\vphantom{13pt}\vphantom{13pt}\vphantom{13pt}\vphantom{13pt}\vphantom{13pt}\vphantom{13pt}\vphantom{13pt}\vphantom{13pt}\vphantom{13pt}\vphantom{13pt}\vphantom{13pt}\vphantom{13pt}\vphantom{13pt}\vphantom{13pt}\vphantom{13pt}
\caption{Observational details.}
\centering
\begin{tabular}{llllllllllllllllll}
\hline
object name       & a $^{(1)}$    & e $^{(2)}$    & i $^{(3)}$     & H $^{(4)}$     & Date              & R $\mathrm{_H}$$^{(5)}$ & $\Delta$ $^{(6)}$ & $\alpha$ $^{(7)}$     & $\theta_s$ $^{(8)}$                                                     & $\chi_{\mathrm{am}}$ $^{(9)}$ & g,r,i obs $^{(10)}$ & $\Delta$T $^{(11)}$ & $\delta\dot{\theta}$ $^{(12)}$  & $\theta_t$ $^{(13)}$                                               & N$_{\theta_t}$ $^{(14)}$ & $\delta t$ $^{(15)}$ & m$\mathrm{_r}$ $^{(16)}$ \\
               & (au) &      & (deg) & (mag) & UTC               & (au)           & (au)     & ($^{\circ}$) & \begin{tabular}[c]{@{}l@{}}($^{\prime\prime}$)\end{tabular} &             &           & (s)       & ($^{\prime\prime}/s$) & \begin{tabular}[c]{@{}l@{}}($^{\prime\prime}$)\end{tabular} &                & (s)        &        (mag)        \\ \hline
2016 CG$_{18}$ & 1.06 & 0.12 & 5.31  & 28.5  & 2016 Feb 05-13:52 & 0.989          & 0.004    & 45.4         & 0.56                                                           & 1.14        & 2,2,2     & 749.6     & 0.55                  & 33.1                                                           & 59.1           & 1.01       & 17.29$\pm$0.01 \\
2016 EV$_{84}$ & 0.87 & 0.18 & 13.62 & 26.7  & 2016 Mar 12-09:17 & 1.011          & 0.018    & 14.4         & 0.71                                                           & 1.03        & 2,2,1     & 754.8     & 0.28                  & 16.8                                                           & 23.7           & 2.53       & 18.89$\pm$0.01 \\
2016 GE$_{1}$  & 2.10 & 0.53 & 10.77 & 26.7  & 2016 Apr 04-10:07 & 1.007          & 0.009    & 41.8         & 0.67                                                           & 1.01        & 2,3,2     & 935.8     & 0.32                  & 19.3                                                           & 28.8           & 2.08       & 17.34$\pm$0.01 \\ \hline
\end{tabular}
\begin{tablenotes}
\item \textbf{Notes.} (1) semi-major axis, (2) eccentricity, (3) inclination, (4) absolute magnitude, (5) heliocentric distance, (6) geocentric distance, (7) phase angle, (8) seeing FWHM measured using image background stars, (9) airmass measured at chip-center, (10) number of g, r and i exposures, (11) duration of entire g, r and i sequence, (12) net difference between target movement rate and the tracking rate of the telescope, (13) asteroid trail length, (14) number of resolution elements in asteroid trail, (15) time resolution element per unit of spatial resolution along asteroid trail, (16) apparent r-band magnitude.
\end{tablenotes}
\end{table}
\end{landscape}


\bsp	
\label{lastpage}
\end{document}